\documentclass{aa}  

\usepackage{graphicx}
\usepackage{txfonts}

\usepackage{graphicx}	
\usepackage{amsmath}	
\usepackage{amssymb}	
\usepackage[dvipsnames]{xcolor} 
\usepackage{nicefrac,xfrac}
\usepackage{verbatim}
\usepackage{booktabs, threeparttable}

\usepackage{hyperref}
\hypersetup{
    colorlinks=true,
    citecolor=blue,
    linkcolor=blue,
    filecolor=magenta,      
    urlcolor=cyan,
    pdftitle={Overleaf Example},
    pdfpagemode=FullScreen,
    }

\newcommand {\kms} {km~s$^{-1}$}

\newcommand {\Reff} {$R_{\rm e}$}
\newcommand {\sersic} {S\'{e}rsic }

\defcitealias{Forbes17_SLUGGS_kin}{F17}
\defcitealias{Bellstedt2018_EnclosedMassProfile}{B18}
\defcitealias{Poci2017_TotalMassSlopes}{P17}
\defcitealias{Alabi2016_EnclosedMass}{A16}

\begin{document}

   \title{Total mass slopes and enclosed mass  constrained by\\globular cluster system dynamics}

   \author{Tadeja Ver\v{s}i\v{c}
          \inst{1,2} \and Sabine Thater\inst{1} \and Glenn van de Ven\inst{1} \and Laura L. Watkins\inst{3} \and Prashin Jethwa\inst{1} \and Ryan Leaman\inst{1} \and Alice Zocchi\inst{1}
          }

   \institute{Institute for Astronomy (IfA), University of Vienna,
              T\"urkenschanzstrasse 17, A-1180 Vienna, Austria\\
              \email{tadeja.versic@univie.ac.at}
         \and
             European Southern Observatory, Karl-Schwarzschild Strasse 2, 85748 Garching, Germany
         \and
            AURA for the European Space Agency, ESA Office, Space Telescope Science Institute, 3700 San Martin Drive, Baltimore MD 21218, USA
             }

   \date{Received July XX, 2023; accepted September XX, 2023}
   \titlerunning{Galaxy halos with GCS dynamics}
   \authorrunning{Ver\v{s}i\v{c} et al.}

 
  \abstract{ 
  We study the total-mass density profiles of early-type galaxies (ETGs: ellipticals and lenticulars) with globular clusters (GCs) as kinematic tracers.
  The goal of this work is to probe the total mass distribution, parametrized with a double power-law profile, by constraining the parameters of the profile with a flexible modelling approach.
  To that end, we leverage the extended spatial distribution of GCs from the SLUGGS survey ($\langle R_{\rm GC,\ max} \rangle \sim 8R_{\rm e}$) in combination with discrete dynamical modelling.
   We use discrete Jeans anisotropic modelling in cylindrical coordinates to determine the velocity moments at the location of the GCs in our sample. Assuming a Gaussian line-of-sight velocity distribution (LOSVD) and a combination of informative and uninformative priors we use a Bayesian framework to determine the best-fit parameters of the total mass density profile and orbital properties of the GCs.
   We find that the choice of informative priors does not impact the enclosed mass and inner slope measurements.
   Additionally, the orbital properties  (anisotropy and rotation of the dispersion-dominated GC systems) minimally impact the measurements of the inner slope and enclosed mass. 
   A strong presence of dynamically-distinct subpopulations or low numbers of kinematic tracers can bias the results. 
   Owing to the large spatial extent of the tracers our method is sensitive to the intrinsic inner slope of the total mass profile and we find $\overline{\alpha} = -1.88\pm 0.01$ for 12 galaxies with robust measurements.
   To compare our results with literature values we fit a single power-law profile to the resulting total mass density. 
   In the radial range 0.1-4~\Reff\ our measured slope has a value of $\langle \gamma_{\rm tot}\rangle = -2.22\pm0.14$ and is in good agreement with the literature. 
   Due to the increased flexibility in our modelling approach, our measurements exhibit larger uncertainties, thereby limiting our ability to constrain the intrinsic scatter $\sigma_{\gamma}$.
   }


   \maketitle
%

\section{Introduction}

In the current cosmological framework, galaxy formation is driven by the coalescence of gas in the dark matter overdensities of the large-scale structures \citep{White_Reese_1978_Gal_Formation}. 
In the context of $\Lambda$CDM cosmology, the total mass of galaxies is dominated by dark matter on large scales, while the inner regions $r\lesssim1-2$~\Reff\footnote{\Reff\ is the stellar effective radius where half the stellar light is enclosed.} are dominated by stars \citep{Cappellari+13_Atlas3D_dyn}.
Therefore, in order to understand the mass content of these galaxies we need extended kinematic tracers of the gravitational potential.

Early analytic efforts to characterize the mass distribution within ETGs were based on the dynamical formalism of Jeans equations \citep{Jeans_1922} and implementing the effects of violent relaxation \citep{Lynden-Bell67_ViolentRelaxation}.
With additional assumptions of spherical symmetry and isotropic orbital distribution, the resulting radial density profile for both dark matter and baryons was found to have a power-law distribution $\rho (r) \propto r^{\gamma_{\rm tot}}$ with the slope of -2.
A similar "isothermal" distribution was also found for gas-rich late-type galaxies (LTGs) \citep[see review from ][]{Courteau+14_Masses_of_gals_review} and it results in their flattening of the circular rotation curves first reported by \cite{Rubin+70_M31_Rota_Curve}.

First studies with gravitational lensing of ETGs found that the total mass distribution within the Einstein radius ($\sim$\Reff/2) can be well approximated by a single power law with a narrow scatter and a profile that is slightly steeper than an isothermal sphere with a slope of -2 \citep{Auger2010, Barnabe_11_SLAC_innerSlope_DM_fraction, Sonnenfeld+13_gravitational_lensing_Gamma_redshift_evolution}.
This trend was initially determined for the strongest gravitational lenses, i.e. the most massive galaxies. Subsequently, a similar trend was confirmed for lower-mass galaxies using dynamical modelling
\citep[e.g.][hereafter \citetalias{Alabi2016_EnclosedMass, Poci2017_TotalMassSlopes, Bellstedt2018_EnclosedMassProfile}]{Alabi2016_EnclosedMass, Poci2017_TotalMassSlopes, Bellstedt2018_EnclosedMassProfile}.
In the case of ETGs, the apparent tendency that two non-isothermal distributions of stars and dark matter "conspire" in a slope of close to isothermal has been a point of great discussion.
The "bulge-halo" conspiracy was found to hold only on certain mass scales, suggesting galaxy feedback processes might play a role \citep[and references therein]{Dutton+14_explaining_small_Scatter}.

The flat velocity dispersion profile found within 1~\Reff\ of ETGs has been linked to the dry major mergers as the main formation channels of these gas-poor systems.
Through simulated binary mergers, \cite{Remus_13} showed that the inner slope $\gamma_{\rm tot}\sim -2$ acts as a natural attractor and once in place only a major merger with a high gas fraction can modify the total mass density slope.
They found that the larger the fraction of in-situ formed stars the steeper the total density profile. 
They link this with the dissipationless nature of gas, which sinks in the centre of the merged galaxy to form a new population of more centrally concentrated stars, thereby steepening the total mass slope.

\citetalias{Poci2017_TotalMassSlopes} investigated the correlations between the inner slope and stellar kinematic properties and found that for galaxies with $\log\sigma_{\rm e} \gtrsim 2.1$, the stellar velocity dispersion at 1~\Reff, the inner slope is constant. 
Recently,\ \cite{Zhu_K+23_Manga} extended the study to late-type galaxies (LTGs) and found morphological and mass dependence of the $\gamma_{\rm tot}$ in the sample of MANGA galaxies.
LTGs with $\log\sigma_{\rm e} < 2.2$ show an increasingly shallower total mass-density profile in contrast with ETGs.
They also found a steeper slope with increasing age of the stellar component at a fixed value of $\sigma_{\rm e}$.

In LTGs, the cold gas traces the total mass distribution out to few \Reff, while most of ETGs have already depleted their main gas reservoir and the observations of the centrally concentrated inner stellar component are needed to constrain the potential.
The stellar surface brightness profile drops rapidly and it becomes observationally challenging to observe beyond a few \Reff.
However, all galaxies have their inner and extended outer haloes populated by globular clusters (GCs) \citep[and references therein]{Harris91_GCS_review}.
Owing to their compact size and high stellar surface brightness they can be observed in nearby galaxy clusters \citep[e.g.][]{Cote_04_ACSVCS_survey_intro,Acsfcs+07_FCS_} and they probe the gravitational potential in the outer regions of galaxies and clusters \citep[e.g.][]{Avinash+21_FORNAX_GCS}.
Several literature studies have utilized these tracers to carry out dynamical modelling and used the distinct origin and dynamical properties of the GCs populations to constrain the enclosed mass and to break the mass - anisotropy degeneracy \citep[e.g.][]{Napolitano+14_5846, Zhu16_5846}.

In this paper, we carry out discrete anisotropic Jeans modelling accounting for the observed flattening of the GC systems (GCS) to investigate the inner slope of the total mass profile and the virial mass.
In Sect.~\ref{ch:data} we introduce the sample of GCs used throughout the paper and discuss the source of photometry and kinematics.
Then we discuss the modelling of the tracer density profile in Sect.~\ref{ch:TD_profile}, one of the key components of the dynamical modelling. 
We introduce the dynamical modelling approach we adopt, focusing on the Bayesian inference and different modelling set-ups in Sect.~\ref{ch:DM_main}, followed by a discussion of the results and exploration of potential biases in our modelling approach in Sect.~\ref{ch:Results_RunResults}.
In Sect.~\ref{ch:comparison_with_lit} we compare the results with literature values and discuss our results in Sect.~\ref{ch:Discussion}.
Sect.~\ref{ch:Conslussions} presents the  summarises our main conclusions

\section{Sample of globular cluster systems}\label{ch:data}

In this section, we will introduce the data used throughout this work.
For dynamical modelling, we need kinematic data as well as information on the spatial distribution of the tracer population.
We use the sample of the GC radial velocities (RV) from the SAGES Legacy Unifying Globulars and GalaxieS \citep[SLUGGS; ][]{Brodie+14_SLUGGS_survey_description} survey that observed 27 nearby galaxies.
First, we will discuss the properties of the RV sample and define kinematic flags based on the visual inspections of the GC velocities.
Secondly, we will present the sources of GC photometry for 21 galaxies used to determine the tracer density profile in Sect.~\ref{ch:TD_profile}.
Observed properties of the 21 galaxies we modelled in this work are presented in \autoref{table:galInfo}.

\begin{table*}
\caption{Basic information for the 21 galaxies from the SLUGGS survey modelled in this work. Column (1) gives the NGC identifier of each galaxy.
Columns (2)-(7): distance, stellar mass, stellar effective radius in angular and physical units, environment (F - field, G - group, C - cluster), systemic velocity of the galaxy, all taken from \cite{Forbes17_SLUGGS_kin} and references therein.
Columns (8) and (9) show the inclination (sources listed below) and the number of GCs used in the dynamical modelling, respectively. Column (10) gives the kinematic quality flags discussed in Sect.~\ref{ch:kin_flags}. In column (11) we list the source of GC photometry, with the symbol $^{a)}$ we highlight the galaxies for which only the tracer density profile is available in the literature.}
\label{table:galInfo}

\centering
\begin{tabular}{ccccccccccl}
\hline \hline
NGC & $d$ & $\log_{10}M_*$ & \Reff & \Reff & env & $v_{\rm sys}$ & $i$ & $N_{\mathrm{GC}}$& kin. flags & Photometry source  \\
 & $\mathrm{Mpc}$ & $\mathrm{M_{\odot}}$ & arcsec &  $\mathrm{kpc}$ & $\mathrm{}$ & $\mathrm{km\,s^{-1}}$ & $\mathrm{{}^{\circ}}$ & $\mathrm{}$ & & \\
 (1) & (2) & (3) & (4) & (5) & (6) & (7) & (8) & (9) & (10) & (11) \\
 \hline
720 & 26.9 & 11.27 & 29.1 & 3.8 & F & 1745 & $90.0^\ddagger$ & 65 & Good & SLUGGS \citep{Kartha2014_photoNGC720}$^{a)}$ \\
821 & 23.4 & 11.0 & 43.2 & 4.9 & F & 1718 & $75.0^\dagger$ & 58 & Good w. rot. & SLUGGS \citep{Forbes17_SLUGGS_kin}\\
1023 & 11.1 & 10.99 & 48.0 & 2.6 & G & 602 & $74.0^\ddagger$ & 113 & Good w. rot. & \cite{Forbes+14_photo_NGC1023} \\
1400 & 26.8 & 11.08 & 25.6 & 3.3 & G & 558 & $32.5^\star$ & 69 & Pecul. & SLUGGS \citep{Forbes17_SLUGGS_kin}\\
1407 & 26.8 & 11.6 & 93.4 & 12.1 & G & 1779 & $30.3^\star$ & 356 & Subpop. & SLUGGS \citep{Forbes17_SLUGGS_kin}\\
2768 & 21.8 & 11.21 & 60.3 & 6.4 & G & 1353 & $90.0^\dagger$ & 106 & Good & SLUGGS \citep{Forbes17_SLUGGS_kin} \\
3115 & 9.4 & 10.93 & 36.5 & 1.7 & F & 663 & $90.0^\dagger$ & 149 & Good w. rot. &  SLUGGS \citep{Forbes17_SLUGGS_kin} \\
3377 & 10.9 & 10.5 & 45.4 & 2.4 & G & 690 & $89.0^\dagger$ & 122 & Good & SLUGGS \citep{Forbes17_SLUGGS_kin} \\
3607 & 22.2 & 11.39 & 48.2 & 5.2 & G & 942 & $46.0^\ddagger$ & 39 & Good & SLUGGS \citep{Kartha16_LeoIIGroup}$^{a)}$\\
3608 & 22.3 & 11.03 & 42.9 & 4.6 & G & 1226 & $88.0^\ddagger$ & 29 & Good & SLUGGS \citep{Kartha16_LeoIIGroup}$^{a)}$ \\
4278 & 15.6 & 10.95 & 28.3 & 2.1 & G & 620 & $88.0^\ddagger$ & 256 & Good & SLUGGS \citep{Forbes17_SLUGGS_kin} \\
4365 & 23.1 & 11.51 & 77.8 & 8.7 & G & 1243 & $88.0^\ddagger$ & 251 & Good & SLUGGS \citep{Forbes17_SLUGGS_kin} \\
4374 & 18.5 & 11.51 & 139.0 & 12.5 & C & 1017 & $88.0^\ddagger$ & 41 & Pecul. & \cite{Jordan2009_incompletness} \\
4459 & 16.0 & 10.98 & 48.3 & 3.7 & C & 1192 & $48.0^\ddagger$ & 36 & Pecul.  & \cite{Jordan2009_incompletness} \\
4473 & 15.2 & 10.96 & 30.2 & 2.2 & C & 2260 & $81.0^\dagger$ & 105 & Good  & \cite{Jordan2009_incompletness} \\
4486 & 16.7 & 11.62 & 86.6 & 7.0 & C & 1284 & $84.0^\ddagger$ & 650 & Subpop.  & \cite{Jordan2009_incompletness} \\
4494 & 16.6 & 11.02 & 52.5 & 4.2 & G & 1342 & $86.0^\ddagger$ & 105 & Good  & \cite{Humphrey09_NGC4494_photo} \\
4526 & 16.4 & 11.26 & 32.4 & 2.6 & C & 617 & $77.0^\ddagger$ & 107 & Pecul.  & \cite{Jordan2009_incompletness} \\
4564 & 15.9 & 10.58 & 14.8 & 1.1 & C & 1155 & $76.0^\ddagger$ & 26 & Good w. rot. & \cite{Jordan2009_incompletness} \\
5846 & 24.2 & 11.46 & 89.8 & 10.5 & G & 1712 & $89.0^\ddagger$ & 176 & Subpop. & SLUGGS \citep{Forbes17_SLUGGS_kin} \\
7457 & 12.9 & 10.13 & 34.1 & 2.1 & F & 844 & $74.0^\ddagger$ & 21 & Pecul.  & SLUGGS \citep{Forbes17_SLUGGS_kin} \\
\hline 
\end{tabular}
\tablebib{$^\dagger$ \cite{Bellstedt2018_EnclosedMassProfile}; $^\ddagger$ \cite{Cappellari+13_Atlas3D_dyn}; $^\star$ HyperLeda (\url{http://leda.univ-lyon1.fr/}).}
\end{table*}

\subsection{Kinematic datasets}\label{ch:KinData}

The kinematic tracers from the SLUGGS dataset have a typical spatial extent of $\sim8$~\Reff\ up to $R_{\rm max} \sim 15$~\Reff\ and a median of 80 GCs per galaxy, varying from 20 to 600 GCs per individual galaxy.
The median error of the radial velocities is 10-15~\kms, representing around $10\%$ of the typical velocity dispersion of a GCS for SLUGGS galaxies.
The SLUGGS catalogue excludes foreground stars and background galaxies based on the phase-space diagram \citep{Forbes17_SLUGGS_kin}.
Additionally, ultra-compact dwarfs (UCDs) are flagged and we exclude them from the following analysis. 
UCDs can be misidentified as the bright tail of the GCs due to their compact size and magnitude but they have a different origin, and therefore different spatial distribution and orbital properties \citep[e.g. UCDs in NGC~5128:][]{Dumont22_UCDS_NGC5128}.

\subsubsection*{NGC~5846}\label{ap:remove_photo_contaminats}

\begin{figure}
    \centering
    \includegraphics[width = \columnwidth]{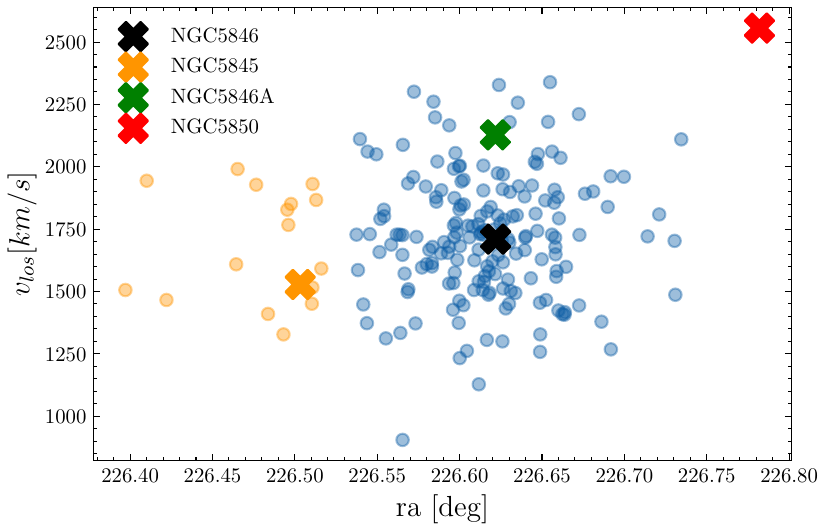}
    \caption{The position of GCs from the SLUGGS survey is shown in blue and orange points.
    The orange points are selected based on their position to be more likely belonging to NGC~5845 and blue points show NGC~5846 GCs used in this work.
    The positions and systemic velocities shown as coloured crosses for the 4 galaxies are taken from Simbad (\url{https://simbad.unistra.fr/simbad/}). }
    \label{fig:NGC5846_contaminants}
\end{figure}

A special remark is needed for galaxy NGC~5846, which is the brightest cluster galaxy and has 3 bright companions (NGC~5846A, NGC~5845 and NGC~5850).
The projected phase space distribution of GCs belonging to NGC~5846 from the SLUGGS survey and the locations of all 4 galaxies are shown in Fig.~\ref{fig:NGC5846_contaminants}.
The right ascension and RV of the GCs are presented with the blue and orange points, the 4 galaxies are presented with coloured crosses.

The spatial and velocity distribution of NGC~5846A and NGC~5845 overlap with those of NGC~5846.
In Fig.~\ref{fig:NGC5846_contaminants} there is a clear elongation of the GCS in the direction of NGC~5845, but the contamination from NGC~5846A is not so apparent in this phase space diagram.
To estimate the contamination of the GC catalogue for NGC~5846 we first estimated the number of GCs for each of these galaxies based on their absolute V-band magnitude.
We used the observed correlations between the number of GCs and the absolute V band magnitude of its host galaxy in a high-density cluster environment \citep[e.g.][]{Zepf+94_SN, Kissler-Patig+97_SpecificFreq_ofGCs}.
We estimate that NGC~5846A hosts fewer than 100 GCs, for NGC~5845 we estimate $\sim300$ GCs and NGC~5846 is expected to host $\sim $2500 GCs.
The GCs from NGC~5845 thus have a much stronger contribution to the contamination than the GCs from NGC~5846A,
To remove potential GCs belonging to NGC~5845 we remove clusters at $\alpha<226.52^{\circ}$ from the kinematic dataset of NGC~5846.
A more sophisticated selection and additional removal of NGC~5846A GCs is beyond the scope of this work.

\subsection{Kinematic flags}\label{ch:kin_flags}

Strong rotation, substructures (non-relaxed populations), and a low number of kinematic tracers or GC subpopulations with distinct kinematics can bias the results of the dynamical modelling. 
Before proceeding with the modelling we visually inspect the velocity distribution of the GCS for all galaxies. 

We focused on the radial variation of mean velocity and velocity dispersion as well as the 2D velocity field of the GCs, like the one shown in the left panel of Fig.~\ref{fig:MagDist} for NGC~3377.
We visually inspected the regularity of the velocity field looking for the presence of strong rotation signatures, cold comoving groups, observationally biased sparse sampling or other peculiar features.

Focusing on these features we defined 4 flags: Good (9/21) and Good w. rot. (4/21) for GCSs showing a highly regular velocity field and either being dispersion-dominated or showing non-negligible rotation respectively.
Galaxies flagged as Subpop. (subpopulations, 3/21) show distinct kinematic subpopulations \citep[such as the blue and red GC populations in NGC~1407,][]{Romanowsky2008_NGC1407} or cold comoving groups that can inflate their velocity dispersion.
Lastly, galaxies flagged as Pecul. (peculiar, 5/21) have on average the lowest number of GCs among the SLUGGSs galaxies and show an irregular velocity field.
Galaxy flags are shown in column 10 of \autoref{table:galInfo}.

\subsection{Photometric datasets}\label{ch:PhotoData}

One of the inputs for the dynamical modelling is the spatial distribution of the kinematic tracers, i.e. tracer density profile, and the photometry is crucial to accurately characterize it.

The SLUGGS catalogue provides photometry in the Sloan g and z bands for 12 galaxies that can be used to construct accurate tracer density profiles.
For the remaining galaxies we utilized either literature photometry, when available, or the parameters of the fitted number density profile. 
With the literature photometric information, we could construct the tracer density profiles for an additional 9 galaxies.
The source of photometry or tracer density profile for each of the galaxies is presented in the last column of \autoref{table:galInfo}.

Eight galaxies in the SLUGGS survey belong to the Virgo galaxy cluster.
This nearby galaxy cluster ($d \simeq 16.5$~Mpc) has been extensively studied with the HST Advanced Camera for Surveys (ACS) as part of the ACS Virgo Cluster Survey (ACSVCS) \citep{Cote_04_ACSVCS_survey_intro}. 
We used the photometry in the F475W ($\simeq$Sloan g) and F850LP ($\simeq$Sloan z) bands of 6 galaxies in the survey.
ACSVCS targeted only the innermost regions of the galaxies and the spatial extent of their tracers is limited to $R_{\rm max} \sim 100~\arcsec$, which translates to $\sim 3$~ \Reff  \ for a typical galaxy in our sample.
For three of the galaxies that we model, only the best-fitting parameters of the tracer density profile were available.

\section{Modeling tracer density profile}\label{ch:TD_profile}

The key component of any dynamical modelling is an accurate spatial distribution of the tracer population.
For the dynamical modelling approach used in this work, the input tracer density profile is needed in the form of Multi-Gaussian Expansion \citep[MGE, ][]{Emsellem94_MGE} and in this section, we explain how we construct MGEs for the SLUGGS galaxies.

We applied the analysis to the 18  SLUGGS galaxies, for which the photometry of the GCs is available in the literature (see \autoref{table:galInfo}).
The sources of photometry used in this work are heterogeneous and assessing magnitude completeness and observational footprint from the photometric images was not possible.
That is why we use the observed GC luminosity function (GCLF) to asses the magnitude completeness and radial variation of PA$_{\rm GCS}$ (position angle) for spatial completeness.
Both steps are applied only to the photometric dataset as dynamical modelling is less sensitive to incompleteness in the kinematic sample.

ETGs show the presence of multiple populations of GCs \citep[][and references therein]{Jordan+07_ACSVCS_GCLF_spread} found in their colour distribution. 
\cite{Fahrion_kin} studied the kinematic properties of GCs in a sample of ETGs from the Fornax 3D survey \citep{F3D_survey} in a similar mass range as the SLUGGS ETGs. 
Splitting them into red and blue populations, based on their colour, they found that the velocity dispersions of both populations agreed within their measurement uncertainties. 
Building on those findings and the work done by \citetalias{Alabi2016_EnclosedMass}, \cite{posti+21_SLUGGS_modeling} and \cite{Bilek+19_SLUGS_GC_MOND},  we treat the red and blue GCs as one population in our dynamical modelling. 
Notable exceptions, where such treatment breaks down, are the central dominant galaxies NGC1407, NGC4486 and NGC5846.
Such galaxies have been shown to have very different accretion histories than other galaxies \citep[e.g.][]{DeLucia_07_Formation_BCGs, Altamariano_15_Accretion_histories_of_BCGs} and much richer population of GCs \citep{Harris_17_GCs_in_BCGs}. 
As \citet{Avinash+21_FORNAX_GCS} showed in a study of the central galaxy of the Fornax cluster, these galaxies host a population of blue (metal-poor) GCs that trace better the group or cluster potential.
We discuss these three galaxies further in Sect.~\ref{ch:Results_effect_of_subpop}.



For the 3 galaxies with literature studies on the GCs number density profile, we use the parameters of the fit.
In Sect.~\ref{ch:TD_NGC3607_3608} we focus on a close pair of Leo II galaxies, for which the GC density profile was investigated with 2 different methods in \cite{Kartha16_LeoIIGroup}.

\begin{table*}

\caption{Results from the analysis of the tracer density profile fitting and completeness.
Column (1) lists the NGC identifier of the galaxy, column (2) the limiting magnitude for completeness as determined using GCLF, (3)  the outer radial limit for spatial completeness as described above in units of \Reff\ of the stellar component of the galaxy. 
Columns (4) and (5) show the best-fit position angle and flattening for the magnitude complete sample of GCS. 
Columns (6) and (7)  show the best-fit effective radius $R_{\rm e, \mathrm{GCS}}$  and \sersic index and $n_{\rm e, \mathrm{GCS}}$ of the number density of the GCS with associated 16th and 84th percentile.
Galaxies NGC~720, NGC~3607 and NGC~3608 have only tracer density profiles presented in the literature and we used those \sersic parameters to perform the  MGE for the dynamical models. 
 } 
\label{table:number_density}

\centering

\begin{tabular}{ccccccc}
\hline \hline
NGC & $mag_{\rm lim}$  & $R_{\rm e,max}$ & PA$_{\mathrm{GCS}}$ & $q_{\mathrm{GCS}}$ & $R_{\rm e, GCS}$ & $n_{\mathrm{GCS}}$\\
 &  & arcsec & $\mathrm{{}^{\circ}}$ &  & arcsec &   \\
(1) & (2) & (3) & (4) & (5) & (6) & (7)    \\
 \hline
$720$ & ---  & --- & 50.0 & 0.77 & $118.2\pm20.4^{a)}$ & $4.16\pm1.21^{a)}$   \\
$821$ & 23.79 & 10.0 &  56 & 0.77 & $147.31_{-34.79}^{+78.24}$ & $2.74_{-1.25}^{+1.47}$   \\
$1023$ & 23.87 & 8.0 &  67 & 0.75 & $104.40_{-7.49}^{+7.57}$ & $1.16_{-0.23}^{+0.33}$   \\
$1400$ & 23.34 & 8.0 & 26 & 0.71 & $247.53_{-88.55}^{+76.70}$ & $3.21_{-1.42}^{+1.21}$   \\
$1407$ & 22.94 & 6.0 & 165 & 0.87 & $276.13_{-53.12}^{+116.57}$ & $2.32_{-0.76}^{+1.20}$   \\
$2768$ & 22.9 & 10.0 & 52 & 0.74 & $92.61_{-19.08}^{+20.89}$ & $2.75_{-0.84}^{+0.83}$   \\
$3115$ & 22.35 & 6.0 & 40 & 0.8 & $214.53_{-85.19}^{+98.07}$ & $3.66_{-1.73}^{+1.54}$   \\
$3377$ & 22.89 & 8.0 &  36 & 0.82 & $160.60_{-40.32}^{+104.70}$ & $3.14_{-1.51}^{+1.95}$   \\
$3607$ & --- & --- & $109\pm8^{b)}$ & $0.61^{b)}$ & $119.4 \pm 17.4 ^{b)}$  & $1.97\pm 1.19^{b)}$    \\
$3608$ & --- & --- & $66\pm 7^{b)}$ & $0.8^{b)}$ & $90 \pm 9^{b)}$  & $0.93\pm  0.56^{b)}$   \\
$4278$ & 22.53 & 12.0 &  168 & 0.93 & $213.84_{-69.11}^{+203.95}$ & $2.53_{-1.12}^{+1.78}$  \\
$4365$ & 22.83 & 10.0 & 39 & 0.73 & $175.68_{-20.36}^{+26.30}$ & $2.45_{-0.65}^{+0.86}$   \\
$4374$ & 25.5 & 1.0 & 140 & 0.89 & $71.69_{-3.60}^{+5.33}$ & $0.69_{-0.13}^{+0.17}$   \\
$4459$ & 24.57 & 2.0 & 46 & 0.85 & $59.45_{-24.48}^{+72.52}$ & $4.13_{-1.93}^{+1.33}$  \\
$4473$ & 24.86 & 3.0 &  110 & 0.82 & $170.63_{-65.77}^{+65.81}$ & $2.67_{-0.92}^{+0.84}$  \\
$4486$ & 25.85 & 1.0 & 19 & 0.94 & $162.46_{-59.52}^{+64.84}$ & $1.17_{-0.39}^{+0.32}$  \\
$4494$ & 23.58 & 10.0 &  10 & 0.75 & $119.95_{-13.71}^{+15.63}$ & $1.90_{-0.58}^{+0.96}$  \\
$4526$ & 24.54 & 3.0 &  128 & 0.74 & $106.35_{-46.06}^{+111.41}$ & $1.64_{-1.03}^{+1.28}$  \\
$4564$ & 24.4 & 8.0 &  48 & 0.67 & $80.98_{-28.44}^{+65.95}$ & $3.07_{-1.06}^{+0.67}$   \\
$5846$ & 22.25 & 10.0 & 12 & 0.87 & $254.20_{-16.18}^{+19.51}$ & $0.49_{-0.14}^{+0.21}$  \\
$7457$ & 22.28 & 6.0 &  131 & 1.0 & $173.58_{-103.14}^{+108.52}$ & $3.58_{-1.89}^{+1.67}$  \\
\hline \hline
\end{tabular}
\tablebib{$^{a)}$   \cite{Kartha2014_photoNGC720}, $^{b)}$ \cite{Kartha16_LeoIIGroup}}
\end{table*}

    \subsection{Magnitude completeness}

We use the observed GCLF to determine the limiting magnitude.
Sources fainter than the limiting magnitude are strongly affected by incompleteness.
Including them in the analysis of the tracer density profile would introduce a bias by mimicking a shallower profile close to the centre of the galaxy as the galaxy light becomes brighter than the faint GCs.

The GCLF for ETGs is well approximated by a Gaussian distribution \citep[][and references therein]{Rejkuba12_GCLF_as_distance_ind_review}.
The mean absolute magnitude of the GCLF, the turn-over magnitude (TOM), is constant for massive ETGs within $\pm0.2$mag \citep[e.g.,][]{Harris+14_universality_of_TOM, Villegas10_GCLF_FCS_VCS} and has even been used to determine precise distances to galaxies \citep[e.g.][]{Jordan+05_TOM_Distance_Indicator}. 
The width of the GCLF, $\sigma_\mathrm{GCLF}$, depends on the total mass of the galaxy with a significant scatter at the faint end \citep{Villegas10_GCLF_FCS_VCS}.

\begin{figure*}
    \centering
    \includegraphics[width=\textwidth]{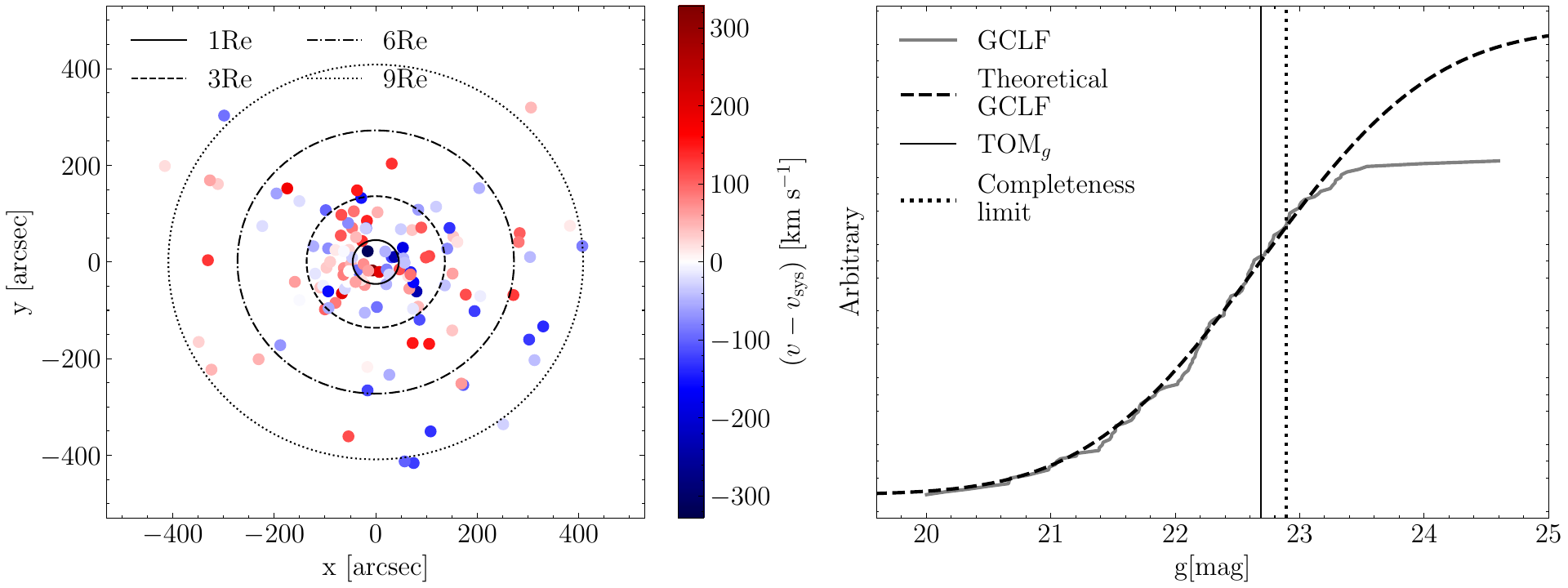}
    \caption{ An example of spatial and magnitude distribution of the kinematic tracers. 
    We used both to determine an observationally unbiased sample of GCs to construct the tracer density profile for each galaxy.
    The left panel shows the spatial distribution of the GCs in NGC~3377 centred at the coordinates of the host galaxy and aligned with the photometric position angle.
    The GCs are colour-coded by their velocity corrected for the systemic velocity of the galaxy. 
    The kinematic tracers extend beyond 9~\Reff.
    The solid grey line in the right panel shows the GCLF of NGC~3377.
    The dashed black curve shows the arbitrarily scaled theoretical GCLF, with the turn-over magnitude marked with a vertical solid black line.
    The dotted vertical line shows the magnitude completeness of the sample, determined as described in the text.}
    
    \label{fig:MagDist}
\end{figure*}

To determine a magnitude-complete sample of GCs, we compare the observed cumulative GCLF with the theoretical Gaussian function, assuming $\mathrm{TOM_V} =$-7.5~mag or $\mathrm{TOM_z} =$-8.5~mag and a spread of $\sigma_\mathrm{GCLF}\sim 1.2$~mag computed from the absolute magnitude of the galaxy in the B band and relation found in \cite{Jordan+07_ACSVCS_GCLF_spread}.
The theoretical luminosity function is normalized to match the number of clusters up to the limiting magnitude in an iterative approach.
For each galaxy, we compute the square of the residuals between the observed cumulative GCLF and theoretical function and we set the magnitude limit where the residuals exceed 1$\sigma$.
This is shown in the right panel of  Fig.~\ref{fig:MagDist} for NGC~3377.
The observed GCLF of this galaxy becomes strongly affected by the incompleteness at the limiting magnitude of 23.8, shown with the grey dotted line in the right panel of Fig.~\ref{fig:MagDist}.
The limiting magnitudes and radial truncation for all of the galaxies are presented in \autoref{table:number_density} together with the results of the \sersic\ fit.

    \subsection{Spatial completeness}

After determining the magnitude limit we analyze the radial variation of  ${\rm  PA}_{\rm GCS}$ (increasing North to East) and flattening defined as $q_{\rm GCS} = 1-e = 1- b/a$ to understand the spatial completeness and measure both parameters.
In the axisymmetric dynamical modelling used in this work, the axis of symmetry has to be aligned with the position angle and can be flattened along the vertical axis.

The analysis is performed in the projected Cartesian coordinate system (x,y), where x is in the direction of West and y North. 
For the transformation from the equatorial coordinates of the GCs $(\alpha_i, \delta_i)$ we use the following equations from \cite{VandeVen06_transform}:
\begin{align}
    \Delta\alpha &= \alpha - \alpha_0 \\
    \Delta\delta &= \delta - \delta_0 \\
    x &= -r_0 \cos(\delta) \sin(\Delta\alpha)\\
    y &= r_0 \big( \sin(\delta) \cos(\delta_0) -
    \cos(\delta) \sin(\delta_0) \cos(\Delta\alpha) \big)
\end{align}
where $\alpha_0, \delta_0$ are the coordinates of the centre of the galaxy. 
The scale factor $r_0 = 180/\pi \times 3600$ converts x and y from radians to arcsec.

The PA$_{\mathrm{GCS}}$ and $q_{\mathrm{GCS}}$ are computed by constructing an inertial tensor from the GC positions (x,y). 
The eigenvectors of the tensor are aligned along the semi-major axes.
We assume that the intrinsic position angle of each GCS is constant with radius and that any deviations we measure are introduced by the observational footprint in the SLUGGS survey.
We take the radius where the PA$_{\rm GCS}$ and $q_{\mathrm{GCS}}$ change rapidly as the maximal completeness radius and we present it in column (3) of \autoref{table:number_density} in units of \Reff.
Columns (4) and (5) show the recovered PA$_{\rm GCS}$ and $q_{\mathrm{GCS}}$ of the spatially and magnitude complete sample.

\subsection{\sersic fits}

After establishing a spatially and magnitude complete sample we could bin the photometric data for each galaxy and fit the MGEs to the binned profile. 
However, binned profiles are stochastic and the number of bins is often small. 
So instead we choose to fit an analytic \sersic function to the binned data as an intermediate step.
To account for the flattening of the GCS we do not bin in spherical radii, instead, we use the elliptical radius for each GC $R_{\rm ell} = \sqrt{x^2 + y^2/q_{\rm GCS}^2}$.

We investigate two binning strategies: logarithmically spaced bins or bins with an equal number of GCs.
For each binning strategy, we additionally create 5 different binned profiles, varying the number of bins or number of GCs per bin.
This step is motivated by the large variation in the number of GCs per galaxy.
It also enables us to compare the resulting best-fit parameters of the analytic function in order to ensure the binning introduces minimal bias.
Then we fit the analytic \sersic\ function to the binned number density $\Sigma\ [N_{\rm GC}\ {\rm arcsec }^{-2}]$
   \begin{equation}
      \nu(R_{\rm ell}) = \nu_0\ \exp\Bigg( -b_n \bigg(\Big(\frac{R_{\rm ell}}{R_e}\Big)^{1/n} - 1 \bigg) \Bigg),\label{eq:Sersic}
   \end{equation}
where $\nu_0$ is the normalisation in the units of $N_{\rm GC}\ {\rm arcsec }^{-2}$, \Reff\ is the effective radius, $n$ is the \sersic index and we use an asymptotic expansion
for $b_n = 2n -1/3 + 4/(405n) + 46/(25515n^2)$ from \cite{Ciotti_Bertin99_Sersic_bn_expansion}.
This approach yields good results even in the presence of low-number statistics with a low number of bins.

To fit the \sersic function we use a Bayesian framework, with flat priors on the free parameters: $n_{\rm GCS}$ and \Reff$_{,\rm GCS}$\ and a Gaussian likelihood.
The best-fit parameters and the posteriors are sampled using \textsc{emcee} \citep{Foreman-Mackey_emcee} a python implementation of the \cite{Goodman_Weare10_affine_sampler} affine-invariant Markov chain Monte Carlo (MCMC) ensemble sampler.

\begin{figure}
    \centering
    \includegraphics[width = \columnwidth]{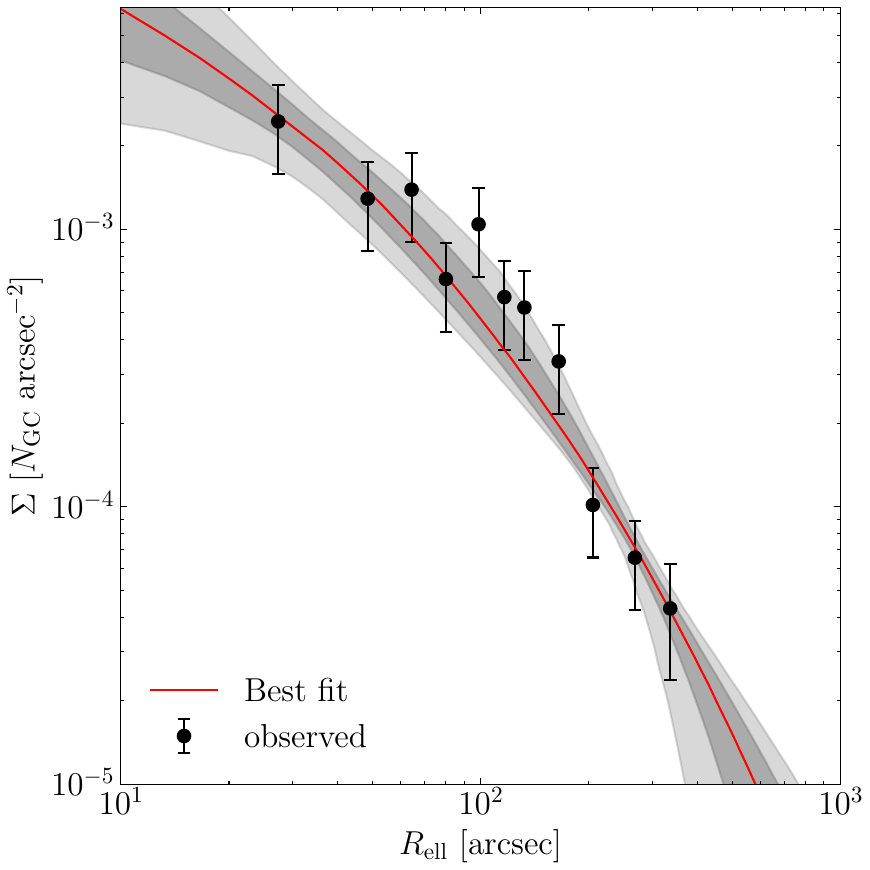}
    \caption{\sersic fit to the binned tracer density profile for NGC~3377 in elliptical radius.
    The binned profile is represented by black points.
    The best-fit model is shown with the solid red line with 1 and 3$\sigma$ uncertainty regions indicated with dark and light grey shaded regions, respectively.
    }
    \label{fig:Sersic}
\end{figure}

As an example, the best-fit \sersic\ model for NGC~3377 is shown in  Fig.~\ref{fig:Sersic}.
The black points with respective uncertainties represent the binned data.
The solid red line is the best-fit model and grey-shaded regions are $1\sigma$ and $3\sigma$ uncertainty regions.
Full results of the fits are presented in  \autoref{table:number_density} in columns (6) and (7).

    \subsection{MGE decomposition}

The last step is to represent the tracer density profile in the MGE format required for the dynamical models we will use later. 
For the 18 galaxies to which we have fit \sersic profiles directly, we draw samples of the \sersic parameters from their posteriors and use them to evaluate the profile at each radius.
This step mimics observational noise and provides a more realistic treatment of the tracer density profile.
We use mgefit, an implementation of the MGE algorithm by \cite{Cappellari02_mgefit}.

Three galaxies, NGC~720, NGC~3607 and NGC~3608, have only parameters of the \sersic profile reported in the literature and do not have GCs photometry available. 
For these we do not have posteriors to draw from, so our approach is necessarily different. 
We generated random samples of $n$ and \Reff\ assuming a Gaussian distribution centred at the best-fit values presented in the literature, generate the perturbed \sersic and proceed in the same way as before.
For these galaxies, we take PA$_{\mathrm{GCS}}$ from the main stellar component and $q_{\rm GCS}$ from the literature studies. 
The information is in \autoref{table:number_density}.


\subsubsection*{NGC~3607 and NGC~3608}\label{ch:TD_NGC3607_3608}

NGC~3607 and NGC~3608 are part of the Leo II group. 
They are a close pair with a projected distance of 39~kpc and a relative difference in the systemic velocity of $\sim$260~\kms \citep{Brodie+14_SLUGGS_survey_description}. 
The velocity dispersion of the GCS, of the order of 150~\kms \citep{Kartha16_LeoIIGroup}, means that membership cannot be assigned based on velocities alone.
To that end, we use the results from \cite{Kartha16_LeoIIGroup}, where the properties of GCS were investigated in a systematic way and two different methods were used to determine the distribution of GCS in both galaxies (see their section 4).

The authors find that PA$_{\mathrm{GCS}}$ and $e_{\mathrm{GCS}}$ for NGC~3607 are consistent between the different methods.
The recovered PA$_{\mathrm{GCS}}$ for NGC~3608 does differ by more than 2$\sigma$ between the two methods.
We adopt the results from their major-axis method (see their section 4.2) for both galaxies and we use the \sersic parameters computed with this method as presented in \autoref{table:number_density}.
We found this method better matched our analysis of the PA$_{\mathrm{GCS}}$ of NGC~3608 based on the kinematic sample.

\section{Discrete dynamical modeling}\label{ch:DM_main}

In this section, we will introduce the dynamical modelling approach used in this work.
First, we will present the anisotropic Jeans models followed by the Bayesian inference to find best-fit parameters.
The results and robustness test against the modelling assumptions are presented in the next section.

\subsection{Axisymmetric Jeans Anisotropic MGE models}\label{ch:Modeling_CJAM}

Most ETGs in the nearby universe are well approximated as axisymmetric systems in a steady state. 
Jeans equations derived from the collisionless Boltzmann equation in cylindrical coordinates  $(R,z,\phi)$ have been used extensively in the literature to determine the dynamical properties and mass profiles of galaxies using a variety of different tracers (PNe, GCs, stellar and gas kinematics).
The recent study of simulated GCS by \cite{Meghan21} investigated the accuracy of Jeans models applied to these discrete potential tracers of LTGs and showed its robustness in recovering the enclosed mass.

The axisymmetric Jeans Anisotropic MGE (JAM) formalism was introduced by \cite{Cappellari2008_JAM}.
Within the formalism, the velocity ellipsoid is cylindrically aligned and velocity anisotropy has the form: $\beta_z = 1-\overline{\varv^2_z}/ \overline{\varv^2_R}$.
To compute the first velocity moments an additional assumption on the rotation of the dynamical tracers is incorporated through the $\kappa$ parameter. 
It encapsulates the relative contributions of ordered and random motions: $\nu \overline{\varv_\phi} = \kappa \Big(\nu\overline{\varv_\phi^2} - \nu\overline{\varv_R^2}\Big)^{1/2}$.
Through the inclination angle, the intrinsic velocity moments are projected along the line of sight.
In this work, we assume the GCS is aligned spatially with the inner stellar component and use the inclination of the latter as presented in the literature from various sources and summarised in \autoref{table:galInfo}.

The tracer and mass densities are represented using the MGE method.
In this method, we represent the number and mass densities as a series of Gaussians.
The benefits of this parametrisation are that it is flexible and computationally fast and stable.

The models provide the first and second velocity moments $\overline{\varv_i},\ \overline{\varv_i^2}$ at a given position on the plane of the sky.
The velocity dispersion is given by: $\sigma_i^2 = \overline{\varv_i^2} - \overline{\varv_i}^2$.
This enables discrete dynamical modelling, through the assumption of local line-of-sight velocity distribution (LOSVD).
There are three main advantages of the discrete approach over binning velocities in radial bins: the number of kinematic data points constraining the model is higher, biases due to an arbitrary binning strategy are removed and contaminants, that could bias binned velocity moments, can be self-consistently modelled \citep[e.g.][]{Watkins2013_CJAM, Henault+18_CJAM_use_example}.
This simplified assumption is reasonable given the quality and size of the dataset and has been used in the literature to model discrete kinematic datasets \cite[e.g.][]{Zhu16_5846}.
\cite{Read2021_ModelingComparison} compared different modelling techniques on simulated datasets (spherical Jeans equations, distribution function models and higher virial moments) and found good agreement in the recovered mass profile for the different modelling techniques.

    \subsection{Gravitational potential}\label{ch:massmodel}

In this work, the total mass in a galaxy (luminous and dark matter) is assumed to follow a 
spherical double power-law profile \citep[as done in ][]{Cappellari+13_Atlas3D_dyn, Cappellari+15_StellarKin_InnerSlope, Poci2017_TotalMassSlopes, Bellstedt2018_EnclosedMassProfile}.
We assumed the following functional form:
   \begin{equation}
      \rho_{\mathrm{tot}}(r) = \rho_s \bigg(\frac{r}{r_s} \bigg)^{\alpha} \bigg( \frac{1}{2} + \frac{r}{2r_s} \bigg)^{-\alpha - 3},\label{eq:NFW_rho}
   \end{equation}
where $\rho_s$ represents scale density and $r_s$ scale radius. The inner total mass slope, parameterized by $\alpha$, transitions to an outer slope fixed at -3.

\subsection{Bayesian inference}\label{ch:bayesian}

We use Bayesian inference to find the best-fit models. 
Each GC is treated independently and no binning in velocities is used to obtain the mean velocity and velocity dispersion profile.

\subsubsection*{Likelihood}

Given the assumption of a Gaussian LOSVD the likelihood is:
   \begin{equation}
      \ln\mathcal{L}_i = - \frac{1}{2}\ln \Big(2\pi \sigma_{los,i}^2\Big) - \frac{(\varv_i - \overline{\varv_i})^2}{2\sigma_{los,i}^2},\label{eq:likelihood}
   \end{equation}
where $\sigma_{los,i}^2 =  \sigma_i^2 + \delta \varv_i ^2 $ is the dispersion of the LOSVD at the position of the $i$th GC convolved with the measurement uncertainty.
At the position of each GC, this constitutes the likelihood of the measured velocity $(\varv_i, \delta \varv_i)$ given the free model parameters $\theta$.

\subsubsection*{Priors and summary of the modelling set-ups}

In this analysis, we use both uninformative, flat priors and informative, Gaussian priors.
The informative priors were necessary to break the degeneracy in the modelled parameters (in particular between $\rho_s$ and $r_s$) without the need for fixing one of them.
The Gaussian function was chosen for its simplicity and symmetry.
To test and ensure that the priors are not driving the resulting posteriors, we run two models with different prior widths.

We consider three different model set-ups: models A, B and C. 
They differ in the number of free parameters and the widths of Gaussian priors.
In models A and B we vary only the three parameters of the total mass profile in Eq.~\ref{eq:NFW_rho}: $(\rho_s, r_s, \alpha)$ and change the prior widths between the two models.
In model C we additionally vary rotation and velocity anisotropy and thus have 5 free parameters: $(\rho_s, r_s, \alpha, \beta_z, \kappa)$.
The modelling setups and choices of priors are summarised in \autoref{table:priors}.

The assumption of no rotation and isotropy in models A and B are motivated by the substantial increase in the computational time of the models when $\beta_z\neq0$ and $\kappa\neq0$.
Before proceeding with models A and B we explored the variation of the likelihood on small grid models by varying $\beta_z$ and $\kappa$.
We found the peak of the likelihood corresponding to very mild radial anisotropy ($\beta_z>0$) and the galaxies which were flagged as having strong rotation signatures in Sect.~\ref{ch:KinData} have a prominent non-zero peak.
We discuss the impact of these assumptions on the recovered parameters of the total mass density in Sect.~\ref{ch:Modeling_effect_of_rotation_anisotropy}.

\begin{table*}
\caption{The overview of the priors for different set-ups. Parameters with uninformative priors ($\alpha$ and $\kappa$) have lower and upper limits (min/max). 
The other three parameters ($\log_{10}\rho_{\rm s}$,
$\log_{10}r_{\rm s}$, and $\beta_{\rm z}$) have Gaussian priors characterized by mean $\mu_\mathrm{prior}$ and spread $\sigma_\mathrm{prior}$.
The models A and B have anisotropy and rotation fixed to 0, these assumptions are relaxed in model C.} \label{table:priors}

\centering
\begin{tabular}{c|cc|cc|cc|cc|cc|l}
\hline \hline

Model &\multicolumn{2}{|c|}
{$\log_{10}\rho_{\rm s}\ [$M$_\odot$ pc$^{-3}]$} & \multicolumn{2}{|c|}
{$\log_{10}r_{\rm s}[$kpc$]$}& \multicolumn{2}{|c|}
{$\alpha$}& \multicolumn{2}{|c|}
{$\beta_{\rm z}$} & \multicolumn{2}{|c|}
{$\kappa$} & Comments\\
 & $\mu_\mathrm{prior}$ & $\sigma_\mathrm{prior}$ & $\mu_\mathrm{prior}$ & $\sigma_\mathrm{prior}$ & min & max & $\mu_\mathrm{prior}$ & $\sigma_\mathrm{prior}$ & min & max \\
 \hline
A & -3 & 1.5 & 1.78 & 1 & -5 & 0 & \multicolumn{2}{c|}{--} & \multicolumn{2}{c|}{--} & Narrow priors\\
B & -3 & 4.5 & 1.78 & 3 & -5 & 0 & \multicolumn{2}{c|}{--} &  \multicolumn{2}{c|}{--} & Wide priors\\
C & -3 & 1.5 & 1.78 & 1 & -5 & 0 & 0.0 & 0.2 & -1 & 1 & Rotating model\\
\hline
\end{tabular}
\end{table*}

The parameters $\rho_s$ and $ r_s$ can vary by several orders of magnitude and to sample the posterior more efficiently they are sampled in logarithmic space.
Other parameters are dimensionless and even though anisotropy is a strongly asymmetric variable\footnote{Velocity anisotropy varies between $1$ and $-\infty$ for extremely radial and vertically anisotropic systems respectively.} we do not reparametrize it as GCSs show only mildly anisotropic orbital distributions as discussed in Sect.~\ref{ch:Modeling_effect_of_rotation_anisotropy}.
The effect our choice of priors has on the posteriors is discussed in Sect.~\ref{ch:R_prior_ef}.

\subsubsection*{Posterior}

Having defined the likelihood and priors, the Bayes theorem gives the expression for the posterior:
   \begin{equation}
      \ln\mathcal{P(\theta|\mathcal{D})} \propto \sum_{i=1}^{N_{GC}} \Big(\ln\mathcal{L}(\mathcal{D}_i| \theta)\Big) + \ln p(\theta) ,\label{eq:posterior}
   \end{equation}
where $\mathcal{P(\theta|\mathcal{D})}$ is the posterior for a given set of free parameters $\theta$, $\mathcal{L}(\mathcal{D}_i| \theta)$ is the likelihood for the $i$th GC and  $p(\theta)$ is the prior.

\subsubsection*{MCMC set-up}

To explore the parameter space and sample the region of best-fitting parameters we use the \textsc{emcee} package, a Python implementation of an affine-invariant ensemble sampler for Markov chain Monte Carlo (MCMC).
For models A and B (isotropic and non-rotating) we use 100 walkers and 6000 steps and for model C (free $\beta_z$ and $\kappa$) 100 walkers and 2000 steps, due to the increase in the computational time.
In both set-ups, 600 steps are discarded as a burn-in phase.
We verified that this setup is longer than several auto-correlation times and we visually inspected the convergence of the chains.
The initial conditions are the same in all runs: $\log_{10}\rho_s = -4\ [\mathrm{M}_\odot \mathrm{pc}^{-3}]$, $\log_{10}r_s = 1.3\ [\mathrm{kpc}]$, $\alpha = -1.5$, for model C: $\beta = 0.01$, $\kappa = 0.1$.
These initial conditions were informed by prior dynamical modelling of GCs in FCC 167, an ETG of the Fornax galaxy cluster.
Given its properties, the modelled galaxy is a good representative of the ETGs in the SLUGGS sample.
We explored the robustness of the posteriors against the variations in the initial conditions and found a negligible effect.

\section{Results: fiducial model and robust sample}\label{ch:Results_RunResults}

\begin{figure}
    \centering
    \includegraphics[width = \columnwidth]{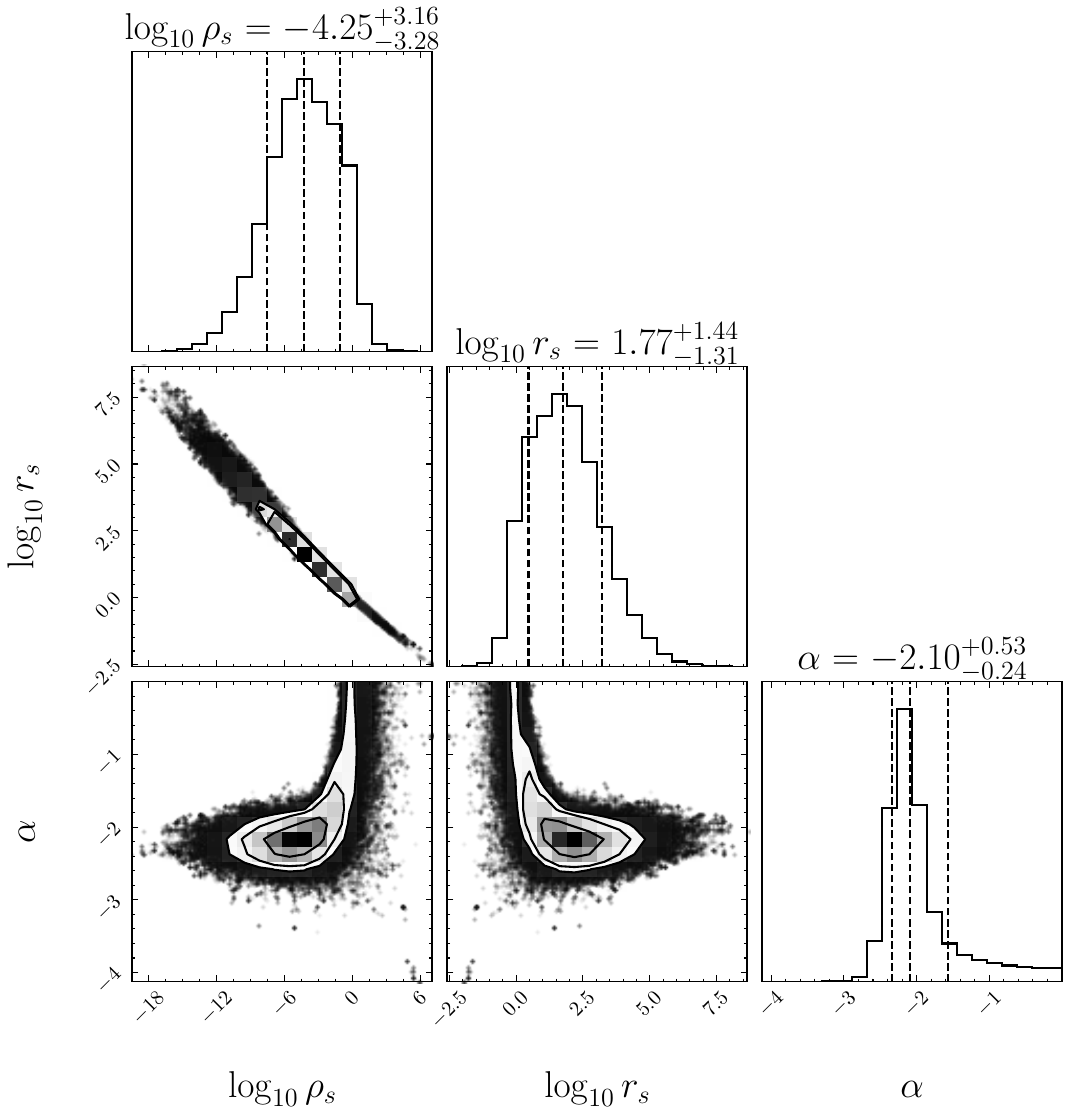}
    \caption{Posterior of model B for NGC~3377. The median and the $1\sigma$ percentiles are shown as dashed lines on each of the marginalized posteriors and the corresponding values are written on top.}
    \label{fig:Fiducial_Corner}
\end{figure}

Now that we have described the dataset and laid out the models we can apply the models to the data.
We applied models A, B and C to all 21 galaxies in our sample and found that all MCMC chains converged and all posteriors were unimodal.
As an example, we show the corner plot for NGC~3377 for the model B setup in  Fig.~\ref{fig:Fiducial_Corner}.
The $\rho_s$ and $r_s$ are strongly anti-correlated and the informative priors we impose help to constrain both but cannot break the degeneracy.
The inner slope $\alpha$ is not degenerate with the scale density and radius but has a tail towards shallower profiles.
At such shallow profiles, $r_s\sim $1~kpc (within the innermost kinematic tracer), therefore in these cases the density is effectively a single power-law profile with the slope of -3 (see Eq.~\ref{eq:NFW_rho}).

We adopt model B as the fiducial model, following the arguments laid out in this section; the robust sample of galaxies are the ones with the flags Good and Good w. rot. as introduced in Sect.~\ref{ch:kin_flags}.
We present the best-fit values of the total mass density profile for the fiducial model in \autoref{table:posteriors} for the 19 galaxies we modelled.
NGC~1400 and NGC~3607 were not modelled due to the literature value of the inclination, which was not consistent with the projected flattening of their GCSs.

\begin{table}
\caption{The summary of the posterior results from the fiducial set-up, model B. We list the median and the 16th and 84th percentile on the parameters of the total mass slope. 
We also report the value of the quality flag in the last column.} \label{table:posteriors}

\centering 
\begin{tabular}{c|cccl}
 \hline\hline
NGC & $\log_{10} \rho_s$ & $\log_{10}r_s$ & $\alpha$ & Flag \\
 & $[\mathrm{M_{\odot}\,pc^{-3}}]$ &  [kpc]  & $\mathrm{}$ & $\mathrm{}$ \\
$(1)$ & $(2)$ & $(3)$ & $(4)$ & $(5)$ \\
 \hline
$720$ & $-0.95_{-4.59}^{+2.58}$ & $0.53_{-0.94}^{+1.39}$ & $-2.36_{-1.24}^{+1.42}$ & Good \\
$821$ & $-1.57_{-4.56}^{+3.08}$ & $0.65_{-1.08}^{+1.49}$ & $-2.46_{-1.26}^{+1.28}$ & Good w. rot. \\
$1023$ & $-3.65_{-3.23}^{+2.39}$ & $1.87_{-1.14}^{+1.72}$ & $-1.64_{-0.34}^{+0.69}$ & Good w. rot. \\
$1407$ & $-5.2_{-1.71}^{+1.17}$ & $4.42_{-1.06}^{+1.57}$ & $-1.06_{-0.09}^{+0.09}$ & Subpop. \\
$2768$ & $-1.09_{-3.18}^{+1.24}$ & $0.66_{-0.54}^{+1.25}$ & $-1.55_{-0.75}^{+1.04}$ & Good \\
$3115$ & $-1.29_{-4.5}^{+2.29}$ & $0.67_{-0.91}^{+1.66}$ & $-2.13_{-0.54}^{+1.21}$ & Good w. rot. \\
$3377$ & $-4.25_{-3.28}^{+3.16}$ & $1.77_{-1.31}^{+1.44}$ & $-2.1_{-0.24}^{+0.53}$ & Good \\
$3608$ & $-2.79_{-3.45}^{+2.73}$ & $1.36_{-1.22}^{+1.94}$ & $-1.61_{-1.34}^{+1.0}$ & Good \\
$4278$ & $-3.0_{-3.7}^{+2.05}$ & $1.54_{-0.9}^{+1.69}$ & $-1.85_{-0.29}^{+0.93}$ & Good \\
$4365$ & $-5.54_{-2.67}^{+2.3}$ & $2.95_{-1.1}^{+1.38}$ & $-1.87_{-0.14}^{+0.2}$ & Good \\
$4374$ & $-3.58_{-2.71}^{+2.07}$ & $2.65_{-1.29}^{+1.85}$ & $-1.29_{-0.47}^{+0.57}$ & Pecul. \\
$4459$ & $-4.27_{-2.59}^{+2.03}$ & $2.57_{-1.27}^{+1.8}$ & $-1.28_{-0.38}^{+0.5}$ & Pecul. \\
$4473$ & $-2.3_{-3.87}^{+1.7}$ & $1.12_{-0.76}^{+1.7}$ & $-1.77_{-0.43}^{+1.05}$ & Good \\
$4486$ & $-4.72_{-2.15}^{+1.52}$ & $3.49_{-1.05}^{+1.67}$ & $-1.21_{-0.1}^{+0.16}$ & Subpop. \\
$4494$ & $-3.84_{-3.4}^{+2.31}$ & $1.73_{-1.06}^{+1.71}$ & $-1.72_{-0.34}^{+0.76}$ & Good \\
$4526$ & $-4.47_{-2.43}^{+2.01}$ & $2.81_{-1.27}^{+1.78}$ & $-1.26_{-0.28}^{+0.41}$ & Pecul. \\
$4564$ & $-2.09_{-4.51}^{+3.53}$ & $0.53_{-1.21}^{+1.59}$ & $-2.51_{-1.32}^{+1.12}$ & Good w. rot. \\
$5846$ & $-4.05_{-1.69}^{+0.97}$ & $3.37_{-1.24}^{+2.02}$ & $-0.61_{-0.43}^{+0.35}$ & Subpop. \\
$7457$ & $-3.51_{-2.92}^{+2.27}$ & $1.6_{-1.29}^{+2.14}$ & $-1.27_{-0.87}^{+0.77}$ & Pecul. \\
 \hline
\end{tabular}

\end{table}

\subsection{Effect of informative priors}\label{ch:R_prior_ef}

\begin{figure}
    \centering
    \includegraphics[width = \columnwidth]{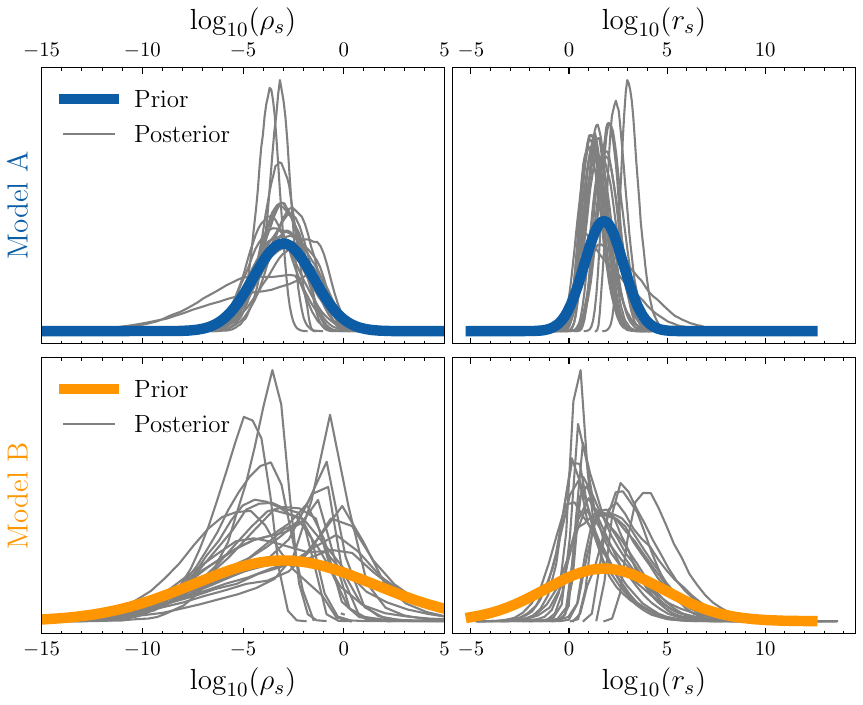}
    \caption{Comparison between the informative priors for $\log_{10}\rho_s$ and $\log_{10}r_s$ in coloured lines for model A and model B and posteriors for each of the 21 modelled galaxies in grey.
    In the fiducial model B the widths of the posteriors are narrower than those of the priors. 
    Which indicates that the data is driving the posterior, not the prior.
    }
    \label{fig:MargenalizedPosteriror_prior}
\end{figure}

To test the robustness of the results against our choice of priors we discuss the marginalized posteriors of models A and B, the two isotropic and non-rotating set-ups, in this section. 
We focus only on the parameters with informative priors $\log_{10}\rho_s$ and $\log_{10}r_s$ as the prior for $\alpha$ is the same in both set-ups (see \autoref{table:priors}).
The posteriors for both set-ups are shown in Fig.~\ref{fig:MargenalizedPosteriror_prior}.
The top row shows the results for model A and the bottom row for model B (fiducial model) as described in Sect.~\ref{ch:bayesian}. 
Each column corresponds to a different free parameter of the total density profile: $\log_{10}\rho_s$ on the left and $\log_{10}r_s$ on the right.
The solid grey lines show the marginalized posteriors for each of the modelled galaxies and priors are shown with coloured lines: the blue line shows prior for model A and the orange for model B.

The posteriors and priors show similar widths on the top panels for model A making it unclear whether the data or prior are driving the posterior distribution in this case.
This motivated us to test the robustness of the informative priors to ensure our choice of prior was not driving the posterior. 
To that end, we increased the widths of the Gaussian priors on the $\log_{10}\rho_s$ and $\log_{10}r_s$ parameters
and on the bottom row for model B the prior is wider than the posteriors.
This indicates the data is driving the posterior for this choice of prior.
Thus we adopted model B as the fiducial setup.
This effect is most strongly seen for the $\log_{10}\rho_s$ parameter, while the posteriors in $\log_{10}r_s$ are narrower than the prior in both instances.
For clarity, the inner slope $\alpha$ is not shown in Fig.~\ref{fig:MargenalizedPosteriror_prior}.
We compared the values of the slope among both setups and found minimal differences in their posteriors.

We did the same investigation for model C, which includes rotation and anisotropy and found similar behaviour of the posteriors as for model A.

We further explored the impact of informative priors by comparing the resulting enclosed mass for models A and B.
To compute the enclosed mass we numerically integrate Eq.~\ref{eq:NFW_rho}.
We randomly drew 1000 samples from the posterior, computed the enclosed mass profile and determined the uncertainties at fixed radii: 1~\Reff, 5~\Reff\ and at the radius of the maximal extent of the tracers for literature comparison.
In the integration, we excluded profiles with $\alpha$ steeper than -3. These result in infinite enclosed mass, and with $\lesssim$0.5\% of the chains having such extreme values we considered this selection did not impact our results and conclusions.

\begin{figure}
    \centering
    \includegraphics[width = \columnwidth]{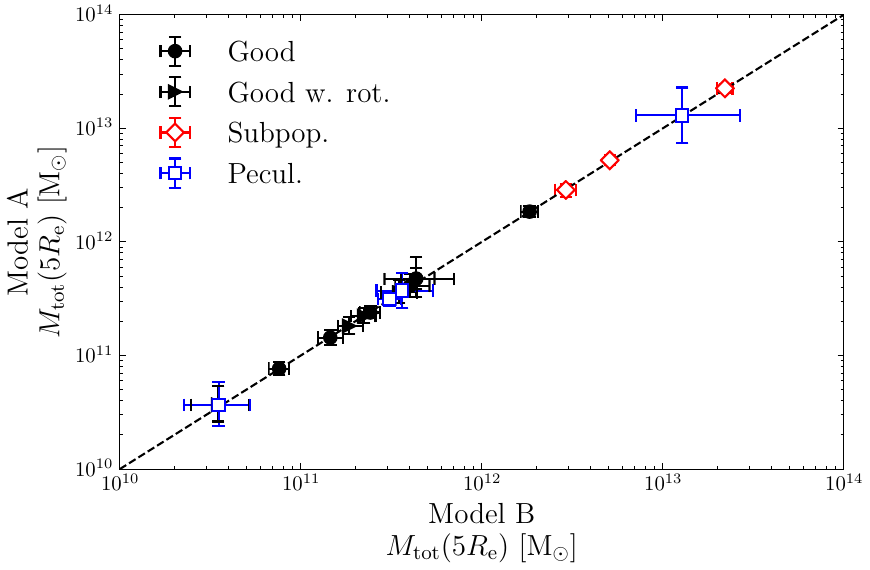}
    \caption{The effect of informative priors on the enclosed mass within 5~\Reff. The black dots and triangles show galaxies flagged as Good and Good w. rot., respectively as defined in Sect.~\ref{ch:kin_flags}. The measurements for these galaxies are robust. Open red diamonds and blue squares highlight galaxies flagged as Subpop. and Pecul. respectively and for which the measurements are not robust.
    The dashed line is a 1-1 line for reference. 
    }
    \label{fig:EnclosedMass_prior}
\end{figure}

The enclosed mass measurements for model B are presented in \autoref{tab:mass_results} and the comparison between the fiducial and narrow set-up is presented in Fig.~\ref{fig:EnclosedMass_prior} within 5~\Reff.
Different symbols correspond with the kinematic flags defined in Sect.~\ref{ch:kin_flags}.
The black points show galaxies flagged as Good or with strong rotation signatures, red open symbols denote the galaxies with strong subpopulations and blue open symbols peculiar galaxies with apparent peculiar velocity fields.
The dashed line shows the 1-1 relation to guide the eye and is not a fit to the points.
This figure demonstrates the negligible effect the prior widths in models A and B have on the enclosed mass.
The same is found for enclosed mass at all radii we investigated, with the uncertainties increasing with increasing radius, and it also holds for model C.

\subsection{Effects of rotation and velocity anisotropy
}\label{ch:Modeling_effect_of_rotation_anisotropy}

After investigating the effects of informative priors on the posterior and enclosed mass we discuss the impact of anisotropy and rotation of the GCS on the parameters of the total mass distribution.
We find the enclosed mass and inner slopes $\alpha$ are consistent between models A, B and C, suggesting that velocity anisotropy and rotation have a negligible impact on the recovered properties of the total mass profile.
\citetalias{Alabi2016_EnclosedMass} studied the GCs of the SLUGGS galaxies and using a Tracer Mass Estimator \citep[TME, e.g.][]{Watkins2010_TME}\footnote{The method assumes spherically-symmetric tracer and mass distributions to evaluate pressure-supported mass estimates and, through additional correction, provides mass corrections due to flattening of the tracer distribution, rotation and velocity anisotropy.} they found minimal contribution ($\sim 6\%$) to the pressure-supported mass estimate due to rotation (see their Figure~7) and $\sim  5\%$ when different velocity anisotropy for GCs is assumed (see the right panel of their Figure~5).  

\begin{figure}
    \centering
    \includegraphics[width=\columnwidth]{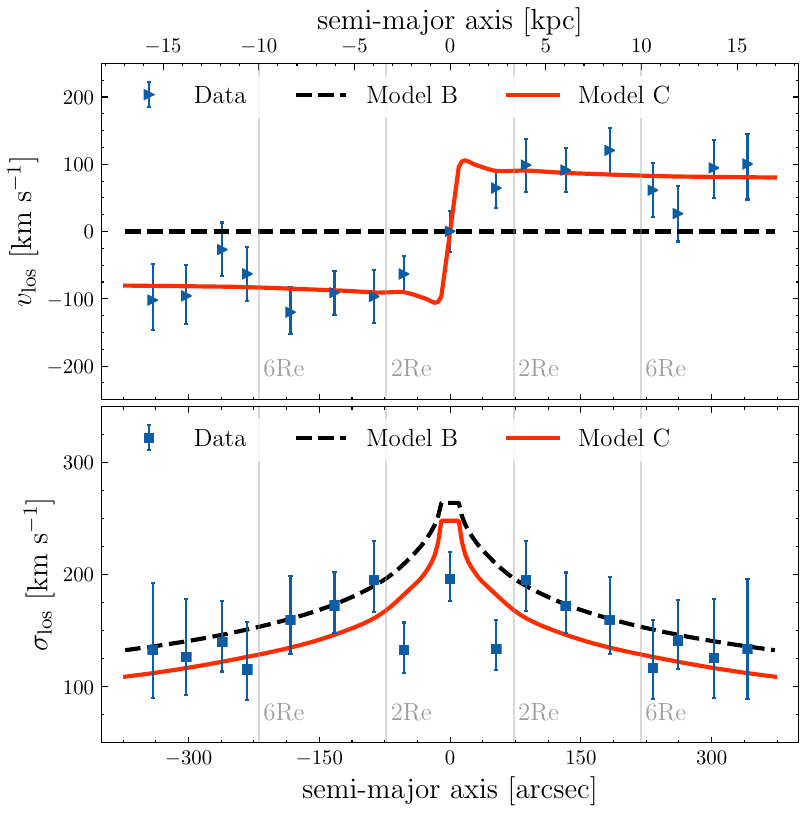}
    \caption{Comparison between the best-fit results from models B and C and the data for NGC~3115  along the semi-major axis of the galaxy. 
    The top panel shows the mean velocity and the bottom the velocity dispersion.
    The blue triangles and squares show the mean velocity and velocity dispersion of the GCs, respectively. 
    The black dashed line shows velocity dispersion computed with the best-fit parameters of model B.
    The solid red line shows the same for model C.
    The vertical grey lines show multiples of the stellar effective radius for comparison.}
    \label{fig:NGC3115_fiducial_and_rotating_model}
\end{figure}

This can be seen in Fig.~\ref{fig:NGC3115_fiducial_and_rotating_model} for NGC~3115; we chose to show this galaxy as it shows signs of strong rotation in the kinematics of its GCs.
In the figure, we compare the velocity moments computed from the best-fit parameters of models B (in black) and C (in red).
Blue squares and triangles show the binned velocity dispersion and rotation of the GCs, they are shown for visualisation purposes and were not used in the dynamical modelling.
To obtain the binned profiles we assumed a point-symmetric velocity field, characteristic of a relaxed axisymmetric system, and symmetrised the velocities to reduce the noise in each radial bin.
We imposed a constant number of GCs in each bin along the semi-major axis and computed robust mean velocity and velocity dispersion, i.e. $\sigma_k^2 = \overline{v_k^2} - \overline{v_k}^2$, for the $k$th bin.
The drop in the velocity dispersion close to the centre of the galaxy is not realistic but is due to incompleteness in radial velocity measurements in the galaxy's inner region.

In this galaxy, the rotation of the GCs in the outer parts is $\overline{v_{\rm los}}\sim100$~\kms\ and velocity dispersion is $\sigma_{\rm los}\sim130$~\kms.
The results for model B are shown with a black dashed line and the velocity moments from \textsc{CJAM} for model C  are shown as a solid red curve. 
In the top panel, we see that model C can reproduce the observed rotation well, which by construction is not the case for model B.
The velocity dispersion profile is very well matched by model B in the bottom panel and model C agrees within the 1$\sigma$ uncertainties of the data.
We can see that in the observing plane, both models B and C agree with the velocity dispersion profile as seen on the bottom panel.

\subsubsection*{Velocity anisotropy and rotation measurements}

\begin{figure}
    \centering
    \includegraphics[width=\columnwidth]{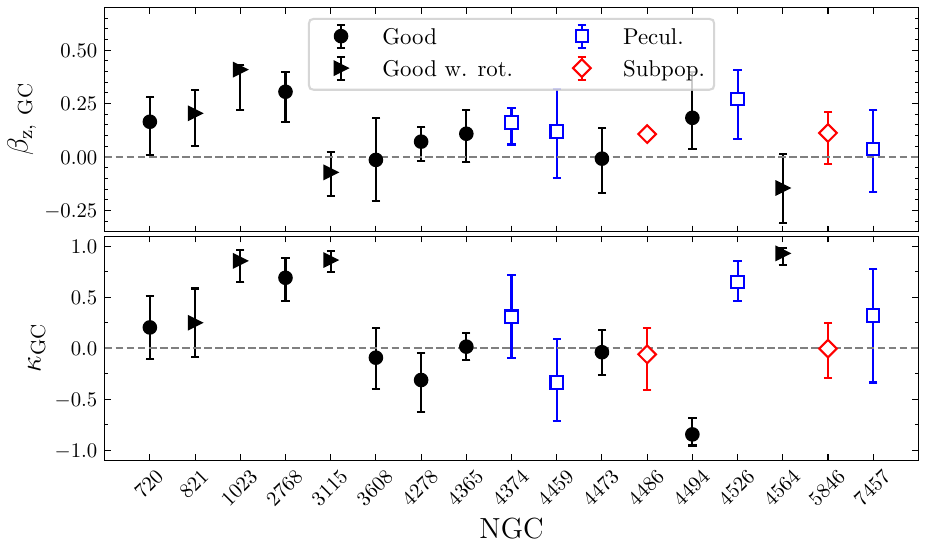}
    \caption{The best-fit velocity anisotropy and rotation from model C. The symbols match those in Fig.~\ref{fig:EnclosedMass_prior}, with black points highlighting the galaxies with robust measurements and with open symbols those without.
    The top panel shows the velocity anisotropy for each galaxy and the bottom panel shows the rotation parameter $\kappa$. 
    The error bars highlight the 16th and 84th percentile uncertainty on the measurement. 
    The dashed grey line shows the isotropic $\beta_{\rm z}=0$ and non-rotating $\kappa=0$ values for reference.}
    \label{fig:velocityAni_rotation}
\end{figure}

In Fig.~\ref{fig:velocityAni_rotation} we show the anisotropy parameter (top panel) and rotation parameter (bottom panel).
We find that most galaxies have velocity anisotropy consistent with zero, consistent with the assumptions in models A and B.
NGC~1023 and 2768 deviate at the level of 3$\sigma$, showing more radial orbits ($\beta_z>0$).
We also find that most GCs are consistent with having negligible rotation, with the exception of those we flagged as having strong rotation signatures, highlighted with black triangles. 
Two galaxies were not flagged as showing a strong rotation signature in the velocity field of their individual GCs (NGC~2768 and NGC~4494), but the model prefers strong $\kappa \neq 0$.

\subsection{Definition of the robust sample}\label{ch:R_Clean_sample}

In this work, we present robust results for 12 galaxies out of the initial 21.
These are flagged as Good (8 galaxies) and Good w. rot. (4 galaxies) in \autoref{tab:mass_results} and in the rest of the paper we refer to these as robust measurements. 

In our analysis, we identified galaxies with strong rotation signatures, low number of GCs or strong presence of subpopulation in their GCS.
Three galaxies, NGC~1407, NGC~4486 (M87) and NGC~5846, are central galaxies of their respective groups and literature studies have identified a strong presence of dynamically distinct populations.
    We flagged them as Subpop. (Subpopulations) based on the differences in the kinematic properties between the red and blue GCs.
We flag four galaxies, NGC~4374, NGC~4459, NGC~4526 and NGC~7457, 
 as Pecul., owing to the peculiar signatures in the kinematics of their GCs.
We will discuss individual galaxies with these flags in more detail in Sect.~\ref{ch:Subpop_Other_lit_comparison}.
In the analysis that follows we show the results for these two classes with open symbols for completeness.

\begin{table}

\caption{Results from the fiducial run (model B set-up). Columns (2) - (3) show the total enclosed mass within specific radii in units of $10^{11}{\rm M}_\odot$. 
The results are given as the median and the uncertainties are computed as $16$th and $84$th percentile
For reference columns (4) and (5) give the effective radii of the galaxy's stellar component in angular and physical units and column (6) includes the quality flags introduced in the text.
Two galaxies (NGC~1400 and NGC~3607) do not have results due to the literature value of the inclination being inconsistent with the flattening of the GCS.
 }\label{tab:mass_results}
 \small
\centering
\begin{tabular}{c|cccc|l}
NGC & $M_{\rm tot}$(1~\Reff) & $M_{\rm tot}$(5~\Reff)  & \Reff & \Reff & Flag \\
 & $[10^{11}{\rm M}_\odot]$ & $[10^{11}{\rm M}_\odot]$  & [arcsec] & [kpc] &  \\
(1) & (2) & (3) & (4) & (5) & (6) \\
\hline \hline
$720$ & $ {1.01}_{-0.8}^{+2.08} $ & $ {3.43}_{-0.7}^{+1.58} $  & 29.1 & 3.8 & Good \\
$821$ & $ {1.0}_{-0.56}^{+1.52} $ & $ {2.81}_{-0.63}^{+1.53} $  & 43.2 & 4.9 & Good w rot \\
$1023$ & $ {0.22}_{-0.09}^{+0.24} $ & $ {1.89}_{-0.32}^{+0.69} $ & 48.0 & 2.58 & Good w rot \\
$1407$ & $ {8.76}_{-1.15}^{+2.31} $ & $ {202.53}_{-20.16}^{+40.28} $ & 93.4 & 12.14 & Subpop. \\
$2768$ & $ {1.15}_{-0.33}^{+0.79} $ & $ {3.66}_{-0.72}^{+1.89} $ & 60.3 & 6.37 & Good \\
$3115$ & $ {0.37}_{-0.26}^{+0.73} $ & $ {1.63}_{-0.24}^{+0.51} $ & 36.5 & 1.66 & Good w rot \\
$3377$ & $ {0.15}_{-0.06}^{+0.14} $ & $ {0.67}_{-0.1}^{+0.2} $ & 45.4 & 2.4 & Good \\
$3608$ & $ {0.3}_{-0.49}^{+1.48} $ & $ {3.05}_{-1.31}^{+4.27} $ & 42.9 & 4.64 & Good \\
$4278$ & $ {0.3}_{-0.15}^{+0.36} $ & $ {2.13}_{-0.28}^{+0.65} $ & 28.3 & 2.14 & Good \\
$4365$ & $ {2.71}_{-0.5}^{+1.07} $ & $ {16.46}_{-1.9}^{+3.96} $ & 77.8 & 8.71 & Good \\
$4374$ & $ {10.47}_{-3.24}^{+7.74} $ & $ {72.89}_{-56.83}^{+185.51} $ & 139.0 & 12.47 & Pecul. \\
$4459$ & $ {0.2}_{-0.09}^{+0.23} $ & $ {2.58}_{-1.09}^{+2.76} $ & 48.3 & 3.75 & Pecul. \\
$4473$ & $ {0.19}_{-0.12}^{+0.29} $ & $ {1.24}_{-0.22}^{+0.47} $ & 30.2 & 2.23 & Good \\
$4486$ & $ {2.41}_{-0.49}^{+0.96} $ & $ {47.78}_{-3.05}^{+6.66} $ & 86.6 & 7.01 & Subpop. \\
$4494$ & $ {0.18}_{-0.07}^{+0.17} $ & $ {1.15}_{-0.23}^{+0.51} $ & 52.5 & 4.23 & Good \\
$4526$ & $ {0.13}_{-0.08}^{+0.19} $ & $ {2.68}_{-0.39}^{+0.89} $ & 32.4 & 2.58 & Pecul. \\
$4564$ & $ {0.07}_{-0.07}^{+0.2} $ & $ {0.26}_{-0.09}^{+0.27} $ & 14.8 & 1.14 & Good w rot \\
$5846$ & $ {0.52}_{-0.3}^{+0.9} $ & $ {25.4}_{-3.66}^{+7.85} $ & 89.8 & 10.54 & Subpop. \\
$7457$ & $ {0.02}_{-0.02}^{+0.08} $ & $ {0.23}_{-0.11}^{+0.31} $ & 34.1 & 2.13 & Pecul. \\
\end{tabular}

\end{table}

\subsection{Single power law slope of the total mass density}\label{ch:LC_SinglePowerLaw_method}

Here we will introduce the method to determine the single power-law slope from the Eq.~\ref{eq:NFW_rho} and present the results in \autoref{tab:Slope_PL}.
The key advantage of this work is the large spatial distribution of the tracers, which enables us to place robust constraints on the total mass profile.
To compare the inner mass slope obtained in this work with literature studies, we fit a single power law with slope $\gamma_{tot}$ in different radial ranges to our best-fit models, and present the results in \autoref{tab:Slope_PL}.
The radial ranges were selected based on literature measurements.

To determine the single power-law slope, we first randomly draw 10000 samples of the total mass parameters from the post-burn-in posterior of each galaxy.
This represents $25\%$ of the total posterior sample and changing the number of samples negligibly affected the final results.
For each set of parameters we solve for the single power-law slope by computing the least square solution of equation: $\log\rho = \gamma_\mathrm{tot} \log r + A$, with the logarithmic slope $\gamma_\mathrm{tot}$ and scaling $A$.
The median and 16th and 84th uncertainty intervals for $\gamma_\mathrm{tot}$ are then computed and presented in  \autoref{tab:Slope_PL} for the 5 different radial ranges.

\begin{figure}
    \centering
    \includegraphics[width = \columnwidth]{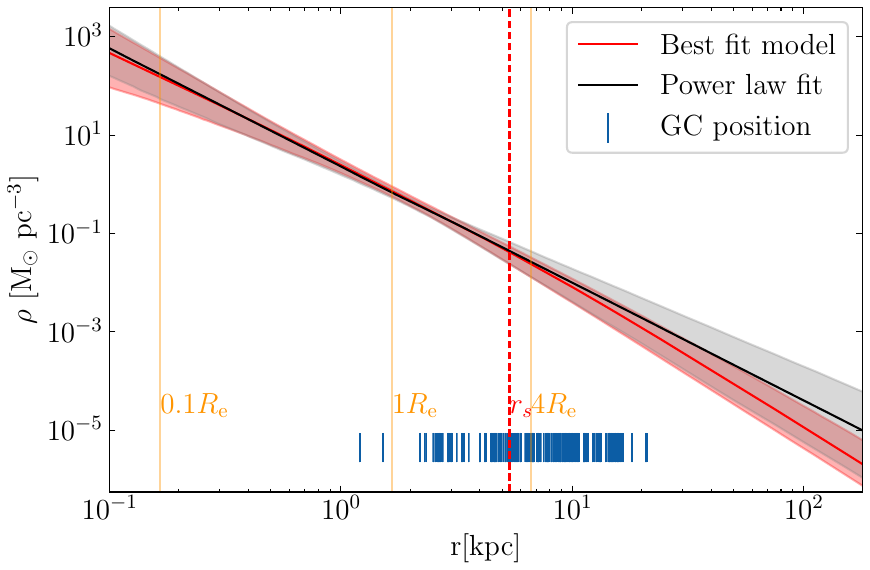}
    \caption{Total mass density of NGC~3115 from the fiducial model. The red solid line with the associated shaded region shows the median, 16th and 84th percentile of the total mass density profile. The red dashed line shows the median scale radius of the total mass density profile.
    The black solid line and grey shaded regions show the resulting single power-law fit in the region $0.1<r/R_e<4$ and 1$\sigma$ uncertainties as explained in Sect.~\ref{ch:LC_SinglePowerLaw_method}.
    Orange vertical lines show the stellar effective radius and blue lines indicate the positions of the tracers for reference. } 
    \label{fig:Total_Mass_distribution_3115}
\end{figure}

To illustrate the method an example of the fit to NGC~3115 is shown in Fig.~\ref{fig:Total_Mass_distribution_3115}.
The red solid line and red shaded regions show the median and $1\sigma$ percentiles of the total density profile.
The vertical red dashed line shows the location of the median scale radius of the total mass density for reference.
For comparison, we chose the $\gamma_{tot}$ and $A$\ obtained from the fit in the range $0.1<r/$\Reff$<4$ (same as \citetalias{Bellstedt2018_EnclosedMassProfile}).
The black solid line shows the power-law with the median and grey-shaded regions highlighting the uncertainty.
The vertical orange lines show the multiples of the \Reff, and the small vertical blue lines indicate the projected positions of the GCs.

Depending on the radial region over which we fit the single power-law profile the resulting power-law slope will vary as also commented on in \citetalias{Bellstedt2018_EnclosedMassProfile}.
The larger the radial extent the more sensitive we are to the transition from the inner slope $\alpha$ to the outer, fixed to $-3$.
While the profiles agree within the 1$\sigma$ uncertainty, it is clear that using the single power-law profile will result in a higher density and larger enclosed mass in the outer regions of the galaxy.

\begin{table*}
\caption{The single power-law slope of total mass density profile in different radial ranges;  kinematic flags are also listed in the last column.}\label{tab:Slope_PL}
\centering
\begin{tabular}{l| ccccc|l}
& \multicolumn{5}{|c}{$\gamma_{tot}$}\\
NGC & $r<1R_e$ & $r<4R_e$ & $0.1R_e<r/R_{\rm e}<1$ & $0.1<r/R_{\rm e}<4$ & $R_{\mathrm{min}}<r<R_{\mathrm{max}}$ & Flag \\
\hline
\hline
$720$ & $-2.56_{-1.0}^{+0.94}$ & $-2.59_{-0.92}^{+0.78}$ & $-2.69_{-0.69}^{+0.67}$ & $-2.76_{-0.53}^{+0.56}$ & $-2.89_{-0.28}^{+0.36}$ & Good \\
$821$ & $-2.57_{-0.98}^{+0.81}$ & $-2.62_{-0.89}^{+0.71}$ & $-2.73_{-0.66}^{+0.61}$ & $-2.8_{-0.51}^{+0.54}$ & $-2.89_{-0.32}^{+0.46}$ & Good w. rot. \\
$1023$ & $-1.64_{-0.32}^{+0.56}$ & $-1.67_{-0.3}^{+0.44}$ & $-1.69_{-0.29}^{+0.39}$ & $-1.74_{-0.27}^{+0.29}$ & $-1.91_{-0.28}^{+0.28}$ & Good w. rot. \\
$1407$ & $-1.06_{-0.08}^{+0.1}$ & $-1.06_{-0.08}^{+0.09}$ & $-1.06_{-0.08}^{+0.09}$ & $-1.07_{-0.08}^{+0.09}$ & $-1.07_{-0.08}^{+0.09}$ & Subpop. \\
$2768$ & $-1.87_{-0.51}^{+0.72}$ & $-1.96_{-0.45}^{+0.56}$ & $-2.14_{-0.4}^{+0.38}$ & $-2.28_{-0.35}^{+0.28}$ & $-2.47_{-0.28}^{+0.27}$ & Good \\
$3115$ & $-2.21_{-0.51}^{+0.78}$ & $-2.25_{-0.48}^{+0.62}$ & $-2.3_{-0.46}^{+0.54}$ & $-2.36_{-0.43}^{+0.4}$ & $-2.61_{-0.31}^{+0.32}$ & Good w. rot. \\
$3377$ & $-2.11_{-0.24}^{+0.37}$ & $-2.12_{-0.23}^{+0.3}$ & $-2.13_{-0.23}^{+0.26}$ & $-2.17_{-0.23}^{+0.21}$ & $-2.28_{-0.23}^{+0.19}$ & Good \\
$3608$ & $-1.84_{-1.15}^{+0.89}$ & $-1.89_{-1.1}^{+0.84}$ & $-2.0_{-0.99}^{+0.9}$ & $-2.11_{-0.89}^{+0.92}$ & $-2.21_{-0.79}^{+0.91}$ & Good \\
$4278$ & $-1.87_{-0.25}^{+0.72}$ & $-1.88_{-0.25}^{+0.59}$ & $-1.89_{-0.24}^{+0.55}$ & $-1.93_{-0.21}^{+0.38}$ & $-2.12_{-0.18}^{+0.19}$ & Good \\
$4365$ & $-1.87_{-0.14}^{+0.2}$ & $-1.88_{-0.14}^{+0.19}$ & $-1.88_{-0.14}^{+0.18}$ & $-1.9_{-0.13}^{+0.14}$ & $-1.92_{-0.13}^{+0.13}$ & Good \\
$4374$ & $-1.32_{-0.43}^{+0.45}$ & $-1.35_{-0.43}^{+0.43}$ & $-1.43_{-0.41}^{+0.45}$ & $-1.51_{-0.42}^{+0.43}$ & $-1.57_{-0.48}^{+0.45}$ & Pecul. \\
$4459$ & $-1.3_{-0.37}^{+0.42}$ & $-1.32_{-0.36}^{+0.4}$ & $-1.34_{-0.34}^{+0.41}$ & $-1.38_{-0.34}^{+0.37}$ & $-1.44_{-0.35}^{+0.36}$ & Pecul. \\
$4473$ & $-1.78_{-0.39}^{+0.8}$ & $-1.83_{-0.36}^{+0.65}$ & $-1.86_{-0.34}^{+0.59}$ & $-1.96_{-0.28}^{+0.41}$ & $-2.17_{-0.28}^{+0.23}$ & Good \\
$4486$ & $-1.21_{-0.1}^{+0.16}$ & $-1.21_{-0.1}^{+0.15}$ & $-1.21_{-0.1}^{+0.15}$ & $-1.22_{-0.1}^{+0.13}$ & $-1.26_{-0.08}^{+0.08}$ & Subpop. \\
$4494$ & $-1.76_{-0.31}^{+0.62}$ & $-1.77_{-0.31}^{+0.48}$ & $-1.8_{-0.29}^{+0.35}$ & $-1.87_{-0.28}^{+0.29}$ & $-2.02_{-0.3}^{+0.27}$ & Good \\
$4526$ & $-1.28_{-0.28}^{+0.39}$ & $-1.29_{-0.28}^{+0.38}$ & $-1.29_{-0.28}^{+0.37}$ & $-1.3_{-0.27}^{+0.33}$ & $-1.38_{-0.25}^{+0.25}$ & Pecul. \\
$4564$ & $-2.66_{-1.06}^{+0.9}$ & $-2.68_{-0.95}^{+0.8}$ & $-2.74_{-0.87}^{+0.77}$ & $-2.8_{-0.7}^{+0.71}$ & $-2.92_{-0.41}^{+0.59}$ & Good w. rot.\\
$5846$ & $-0.65_{-0.41}^{+0.29}$ & $-0.66_{-0.41}^{+0.28}$ & $-0.68_{-0.4}^{+0.28}$ & $-0.72_{-0.37}^{+0.3}$ & $-0.76_{-0.38}^{+0.32}$ & Subpop. \\
$7457$ & $-1.42_{-0.87}^{+0.7}$ & $-1.46_{-0.88}^{+0.66}$ & $-1.5_{-0.93}^{+0.69}$ & $-1.6_{-0.89}^{+0.71}$ & $-1.68_{-0.93}^{+0.73}$ & Pecul. \\
\end{tabular}

\end{table*}

\section{Comparison with literature}\label{ch:comparison_with_lit}

In this section, we compare the results from this work with the literature measurements obtained with different dynamical modelling methods and kinematic tracers. 
We focus on the comparison of the enclosed mass and the slope of the total mass profile.
We discuss the results for all 19 galaxies we modelled but focus on the sample of 12 galaxies with robust measurements as determined in Sect.~\ref{ch:R_Clean_sample}.

We compare our results with three different studies that have systematically investigated the total mass profile of a large subset of SLUGGS galaxies: \citetalias{Alabi2016_EnclosedMass}, \citetalias{Poci2017_TotalMassSlopes} and \citetalias{Bellstedt2018_EnclosedMassProfile}.
Their overlapping science goals but different kinematic tracers or methodologies make them uniquely suitable for comparison with our work.
\citetalias{Alabi2016_EnclosedMass} determined the enclosed mass with GCs from the SLUGGS survey. 
They used TME which assumes spherically symmetric tracer and mass distribution. 
The inner slope $\gamma_{\rm tot}$ was studied by \citetalias{Poci2017_TotalMassSlopes} and \citetalias{Bellstedt2018_EnclosedMassProfile} with stellar kinematics and axisymmetric JAM.
The former used stellar kinematics from Atlas$^{\rm 3D}$ survey \cite{Cappellari+11_Atlas3D_description} which extends out to 1~\Reff\ and the latter combined the central Atlas$^{\rm 3D}$ kinematics with extended stellar kinematics from the SLUGGS survey out to 4~\Reff.
The summary of the literature studies is presented in \autoref{tab:my_lit_modeling_setup_comparison}.

\begin{table*}[t]

    \caption{The properties of the different dynamical studies from the literature in comparison with this work. }
    \label{tab:my_lit_modeling_setup_comparison}
    \centering
    \begin{threeparttable}
    \begin{tabular}{ p{0.08\linewidth} | p{0.1\linewidth} | p{0.06\linewidth} | p{0.07\linewidth} | p{0.1\linewidth} | p{0.08\linewidth} | p{0.08\linewidth} | p{0.11\linewidth} | p{0.06\linewidth} }
    
         & potential tracers & $R_{\mathrm{max}}[R_e]$ & Modeling tool & symmetry & tracer density & mass density & $r_s$ & $\kappa$ and $\beta$ \\
        \hline \hline
        This work & individual GCs & $4-8$ & CJAM & axisymmetry & \sersic & Eq.~\ref{eq:NFW_rho} & free & free\\
        \citetalias{Alabi2016_EnclosedMass} & ensamble GCs & $4-8$ & TME & spherical & power law & power law & -- & 0\\
        \citetalias{Poci2017_TotalMassSlopes} & unresolved stars & $\sim 1$ & JAM & axisymmetry & SB MGE\tnote{$\dagger$} & Eq.~\ref{eq:NFW_rho} & fixed (20~kpc) & free\\
        \citetalias{Bellstedt2018_EnclosedMassProfile} & unresolved stars & $\sim 4$ & JAM & axisymmetry & SB MGE\tnote{$\dagger$} & Eq.~\ref{eq:NFW_rho} & fixed (20~kpc) & free\\
    \end{tabular}
    \begin{tablenotes}[para, flushleft]\footnotesize
\item[$\dagger$] SB MGE - stellar surface brightness multi Gaussian expansion
\end{tablenotes}
\end{threeparttable}
\end{table*}

\subsection{Enclosed mass}

We compute the enclosed mass at different radii from model B as discussed in Sect.~\ref{ch:Results_RunResults} and compare it to literature results
from \citetalias{Alabi2016_EnclosedMass}.
The study modelled 18/19 galaxies presented in this work and they used GCs with TME.
It assumes that both mass density and the tracer density profile are spherically symmetric with a power-law distribution.
The slopes are not free parameters in their modelling.
Instead, for the GC tracer density slope the authors use literature studies to build an observed correlation between the stellar mass of the host galaxy and the power-law slope of the GCS.
To get the slope of the total mass density the study uses simulated ETGs and observations to determine the correlation between the logarithmic slope and stellar mass of the galaxy.
In their modelling, they find that the steeper the slope of either profile the larger the recovered mass, and their treatment of the tracer density profile could result in biased measurements.

\begin{figure}
    \centering
    \includegraphics[width = \columnwidth]{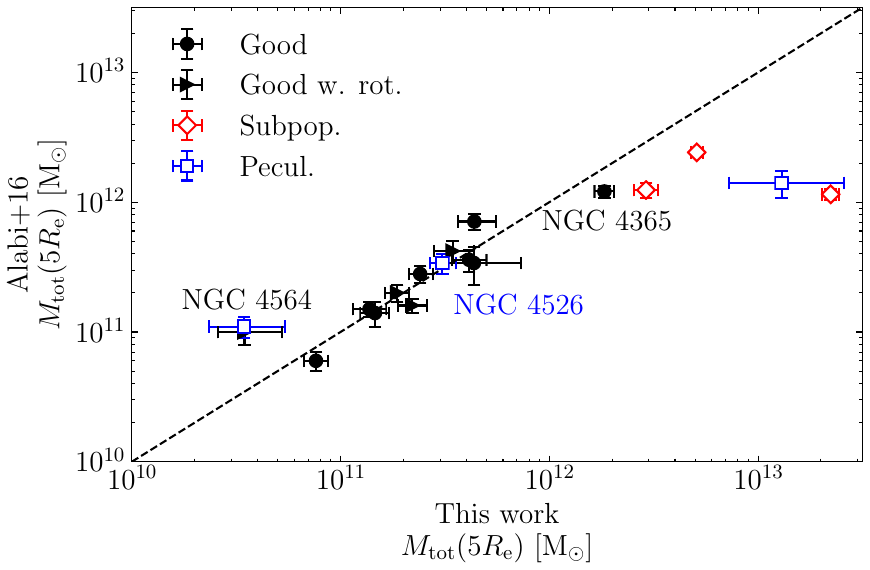}
    \caption{Comparison between the enclosed mass within 5~\Reff\ measured using GC kinematics presented in this work and  \citetalias{Alabi2016_EnclosedMass}. The measurements on the x-axis were obtained with axisymmetric Jeans models assuming general NFW total mass density and on the y-axis the literature study used spherical TME with power-law total mass density distribution.
    The colours and shapes of the points correspond to the ones used in other figures. 
    The dashed line is a 1-1 line for comparison and not a fit to the data. 
    We find consistent enclosed mass for galaxies with robust measurement, with the exception of NGC~4564 and NGC~4365,  highlighted in the figure: both show more than 1$\sigma$ lower and higher values respectively.}
    \label{fig:EnclosedMass_5Re}
\end{figure}

\autoref{fig:EnclosedMass_5Re} shows the enclosed mass within 5~\Reff\ as computed from the fiducial set-up described in Sect.~\ref{ch:Results_RunResults}, on the x-axis, and the values from \citetalias{Alabi2016_EnclosedMass} on the y-axis. 
We use the \citetalias{Alabi2016_EnclosedMass} masses corrected for flattening, rotation and presence of kinematic substructures.
The dashed line shows the 1-1 line as a guide.
Solid black points show galaxies with robust measurements and we highlight the galaxies with strong rotation signatures as triangles.
These galaxies show no systematic difference in the enclosed mass compared with the galaxies that do not show a strong rotation signature.
The biggest outlier is NGC~4564 with mass measurement more than 1$\sigma$ lower than the literature value.
NGC~4365 shows more than 1$\sigma$ higher value of the total mass.

The red diamonds and blue square open symbols indicate the galaxies with strong subpopulations or other complications in the dynamical modelling.
These galaxies deviate most from the 1-1 relation.
Blue points highlight three galaxies: NGC~4374, NGC~4526, and NGC~7457.
\citetalias{Alabi2016_EnclosedMass} found that NGC~4526 and NGC~7457 have a very large number of GCs in kinematic substructures, 25 out of 107 measured GCs and 6 out of 21, respectively. 
We did not remove such kinematic substructures and in the dynamical modelling framework used in this work, the substructures bias the measurements.
NGC~4374 is one of the most massive galaxies in the sample, but it has only 41 observed GCs.
\citetalias{Alabi2016_EnclosedMass} do not find strong corrections due to rotation, flattening or kinematic substructures for this galaxy.
They compare their measurement with the literature and find a range of enclosed mass values, suggesting that this galaxy requires great care when inferring its total mass. 
For our measurement, we conclude that the low number of kinematic tracers is the main driver of the difference with the literature values, but a full investigation is beyond the scope of this work. 
NGC~4526 shows peculiar kinematics but has twice the number of GCs than the other two galaxies.
With 107 GCs Jeans modelling can reasonably constrain the enclosed mass, but the inner slope is a factor of 2 shallower than the literature studies.
Owing to their complex kinematic properties and observational caveats the measurements of these three galaxies are not robust.

\subsection{The power-law slope of the total mass profile}\label{ch:RL_gamma_tot_Lit_compariosn}

To compare the single power-law profiles with the literature we chose \citetalias{Bellstedt2018_EnclosedMassProfile} (their model \textbf{I}) and \citetalias{Poci2017_TotalMassSlopes}  (their model 1).
Both studies used integrated stellar kinematics in different radial ranges.
\citetalias{Poci2017_TotalMassSlopes} used central kinematics out to 0.9~\Reff\ and \citetalias{Bellstedt2018_EnclosedMassProfile} extended the measurements to 4~\Reff.
They both used JAM modelling with cylindrically aligned velocity ellipsoid and the same total mass-density profile to which they fit a single power-law slope.
The spatial extent of their data does not allow to constrain  $r_s$ so they fix it to 20~kpc.
\citetalias{Bellstedt2018_EnclosedMassProfile} explored how their constraints on the $\gamma_{\rm tot}$ change when $r_s$ is left to vary freely.
They claim that in this case, the measurements agree within the 1$\sigma$ uncertainties.
Their similar methodology makes these studies ideal for comparison.
We find that our results agree within the 1$\sigma$ for the robust sample of galaxies, however, our measurements have a larger spread.
We postulate that this could be a consequence of a more flexible modelling approach and fewer data points used to constrain the model.

\begin{figure}
    \centering
    \includegraphics[width = \columnwidth]{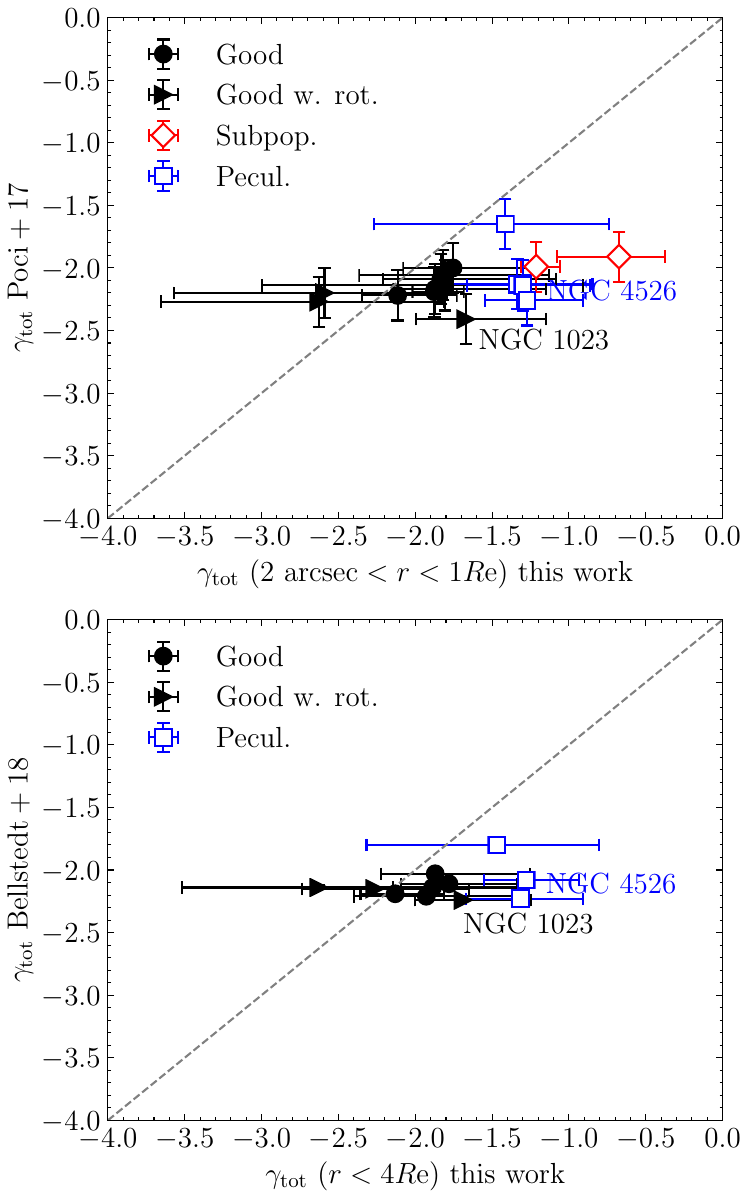}
    \caption{Both panels compare the logarithmic slope of the total mass-density slope presented in this work on the x-axis and the literature study on the y-axis.
    The top panel shows the comparison with the \citetalias{Poci2017_TotalMassSlopes} (model 1) and bottom  \citetalias{Bellstedt2018_EnclosedMassProfile} (model \textbf{I}), the dashed grey line is a 1 to 1 line.
    The colours and symbols match those of Fig.~\ref{fig:EnclosedMass_prior} with solid symbols showing robust measurements and open symbols galaxies excluded from the final analysis. 
    The error bars correspond to the 16th and 84th percentile as presented in \autoref{tab:Slope_PL}.
    In both analyses JAM with integrated stellar kinematics was used, this work uses GC kinematics while literature studies used stellar kinematics.
    Within the uncertainties, the measurements of the robust sample agree with the literature.
    }
    \label{fig:BellSLope}
\end{figure}

The comparison of individual galaxies between this work on the x-axis and literature on the y-axis is shown in Fig.~\ref{fig:BellSLope}.
The top panel shows the comparison with \citetalias{Poci2017_TotalMassSlopes} for the logarithmic slope within 1~\Reff\ and on the bottom panel the comparison with   \citetalias{Bellstedt2018_EnclosedMassProfile} within 4~\Reff.
The symbols correspond to the ones in Fig.~\ref{fig:EnclosedMass_prior}. 
The black points highlight the galaxies with robust measurements and within 1$\sigma$ uncertainties they agree with the literature value. 
NGC~1023 is more than 1$\sigma$ from the dashed 1-1 line and is highlighted in the figure.

Similarly as for the enclosed mass, discussed in Fig.~\ref{fig:EnclosedMass_prior}, the blue open squares and red open diamonds are systematically lower compared with the literature value. 
We highlight galaxy NGC~4526, which shows enclosed mass measurements consistent with the literature in Fig.~\ref{fig:EnclosedMass_5Re} but the inner slope is more than 1$\sigma$ shallower than both literature studies. 
This justifies the exclusion of this galaxy from our robust sample and further supports our initial assignment of kinematic flags.

\subsection{Mean value of the inner slope and the intrinsic scatter}

After determining the samples with robust measurements we investigate the mean value of the inner slope.
We consider both the logarithmic power-law values $\gamma_{\rm tot}$ in different radial ranges and the intrinsic inner slope $\alpha$.
The mean values with respective uncertainties are presented in \autoref{tab:gamma_tot_results} together with a summary of literature results.
The first two rows summarise our results followed by literature values determined using different methodologies and tracers.
The first row summarises the results only for galaxies without strong rotation signatures and the second row includes the galaxies with strong rotation signatures of the GCS.
The third column shows the mean values and 1$\sigma$ error for the intrinsic inner total mass density slope $\alpha$.
The fourth column is the mean value of the single power-law fit within 1~\Reff, for comparison with \citetalias{Poci2017_TotalMassSlopes}, who determined the inner slope of $\langle\gamma_{\rm tot}\rangle_1 = -2.193\pm0.016$.
The last column shows the mean value for the power-law slope within 0.1-4~\Reff, comparable with $\gamma_{\rm tot} = -2.24\pm0.05$ found by \citetalias{Bellstedt2018_EnclosedMassProfile}.

    \begin{table*}
    \caption{Mean values of the inner slope and the observed root-mean-square (rms) scatter for the robust sample of galaxies from this work and literature studies. The first two rows contain results from this work and are followed by 6 literature investigations using varied potential tracers and techniques for z=0 galaxies.
    For the results from this work, the subsamples used to compute the mean values are highlighted in the first column, with the number of galaxies in that subsample in the second column. The third and fourth columns show the statistics of the intrinsic inner slope with 1$\sigma$ uncertainties. The 5th and 6th mean and rms scatter for the logarithmic slope within $0.1-1$~\Reff, and the last two columns show the same for the fit in $0.1-4$~\Reff\ as explained in Sect.~\ref{ch:LC_SinglePowerLaw_method}.
    The last 6 rows contain the values from the literature studies computed in different radial ranges.}\label{tab:gamma_tot_results}
    
    \centering
    \small
        \begin{threeparttable}
    \begin{tabular}{lc|cc|cc|cc}
    Sample & \# gal. & && \multicolumn{2}{c}{$0.1-1$~\Reff} & \multicolumn{2}{|c}{$0.1-4$~\Reff}\\
    & & $\overline{ \alpha}$ & $\sigma_{\alpha}$ & $\langle\gamma_{tot}\rangle_1$ & $\sigma_{\gamma_{tot},1}$ & $\langle\gamma_{tot}\rangle_2$ & $\sigma_{\gamma_{tot},2}$ \\
    \hline \hline

good & 8 & -1.76$\pm$0.01 & 0.83$\pm$0.01 & -2.03$\pm$0.18 & 0.53$\pm$0.19 & -2.12$\pm$0.16 & 0.48$\pm$0.17 \\
good + w. rot   & 12 & -1.88$\pm$0.01 & 0.94$\pm$0.01 & -2.14$\pm$0.16 & 0.64$\pm$0.16 & -2.22$\pm$0.14 & 0.58$\pm$0.15 \\

    \hline
    \cite{Auger2010} & 73 & -- &  --  & $-2.078 \pm 0.027^{1}$ & $0.16\pm0.02^{1}$ & --   & --   \\ 
    \cite{Cappellari+15_StellarKin_InnerSlope} & 14 & -- & -- & $-2.15 \pm 0.03$ & 0.1 & $2.27 \pm 0.06^{2}$ & 0.16$^\dagger$\\ 
    \cite{Serra+16_stellar_gas_innerslope_scatter} & 16 & -- & -- & $-2.18 \pm 0.03^{1}$ & $0.11^{1,}$$^\dagger$ & -- & -- \\    
    \cite{Poci2017_TotalMassSlopes} & 142 & -- & -- & $-2.193 \pm 0.016^{3}$ & 0.15$^\dagger$ & -- & -- \\ 
    \cite{Yildirim+17_total_mass_Slopes} & 16 & -- & -- & -- & -- & $-2.25$ & --  \\ 
    \cite{Bellstedt2018_EnclosedMassProfile} & 21 & -- & -- & $-2.12 \pm 0.05$ & -- & -- & -- \\ 
    \hline
    \end{tabular}
    \begin{tablenotes}[para, flushleft]\footnotesize
\item[1] Single power law profile used in the fit.
\item[2] In radial range $r<4$~\Reff.
\item[3] For model \textbf{I}.
\item[$\dagger$] Estimated intrinsic scatter $\sigma_\gamma$.
\end{tablenotes}
\end{threeparttable}

    \end{table*}

Including only the measured slopes of galaxies flagged as Good results in a slightly shallower slope than when galaxies with strong rotational signatures are included. 
When investigating the effect of rotation and velocity anisotropy we did not find evidence linking rotation to  a steeper profile and we conclude that our kinematic classification is not linked to the measured values.
Our measurements are in broad agreement with these literature studies.
\cite{Auger2010} finds slightly shallower slopes, which could be a result of the systematically more massive galaxies that were probed in this study with gravitational lensing.
\citetalias{Poci2017_TotalMassSlopes} and \cite{Yildirim+17_total_mass_Slopes} use central stellar kinematics, \citetalias{Bellstedt2018_EnclosedMassProfile} and \cite{Cappellari+15_StellarKin_InnerSlope} used extended stellar kinematics and \cite{Serra+16_stellar_gas_innerslope_scatter} combined the central stellar kinematics with extended gas measurements; they all agree within 1$\sigma$ uncertainties.

\begin{figure}
    \centering
    \includegraphics[width = \columnwidth]{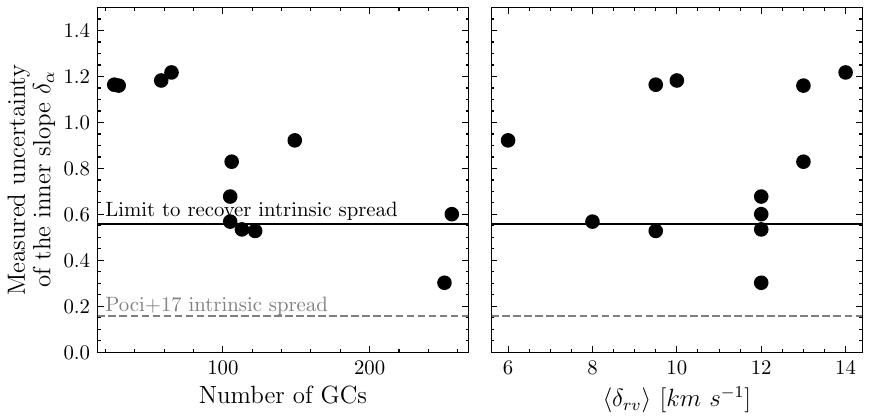}
    \caption{The correlation between the measurement uncertainty of the inner slope (from model B) with the number of GCs per galaxy on the left panel and the median error of the RV on the right panel.
    The grey dashed line indicates the intrinsic scatter of the inner slope of 0.16 as determined by \citetalias{Poci2017_TotalMassSlopes} and \cite{Auger2010}.
    In order to constrain the intrinsic slope on the same level the black solid line indicates the maximal measurement uncertainty needed when using $\sim$10 galaxies.
    }
    \label{fig:IntrinsicSlope}
\end{figure}

The literature studies disagree on the value of the $\sigma_{\gamma_{tot}}$, with \cite{Cappellari+15_StellarKin_InnerSlope} and \cite{Serra+16_stellar_gas_innerslope_scatter} preferring even smaller intrinsic scatter than \cite{Auger2010} and \cite{Serra+16_stellar_gas_innerslope_scatter}.
This low intrinsic spread has been linked with the universality of the inner total mass slope over several magnitudes in stellar mass.
With the novel measurements presented in this work, we attempted to constrain the intrinsic spread of the inner slope $\sigma_\alpha$ \citep[e.g.][eq. 8]{Kirby_11_intrinsic_spread}.
We find that the measurement uncertainties $\delta_\alpha$ are such that the intrinsic spread cannot be recovered.
This is mainly driven by the low average number of GCs per galaxy, which is strongly correlated with the resulting measurement uncertainty as seen on the left panel of Fig.~\ref{fig:IntrinsicSlope}.
The figure shows how the $\delta_\alpha$ for each galaxy is correlated with the number of kinematic tracers on the left panel and the median RV error for GCs on the right panel.
The measurement uncertainty is strongly correlated with the number of kinematic tracers while up to the median RV error of 14~\kms\ the $\delta_\alpha$ is not correlated with the velocity errors.
The dashed grey line indicates the literature value of the $\sigma_{\gamma_{\rm tot}}\sim 0.16$ consistently found by \cite{Auger2010} and \citetalias{Poci2017_TotalMassSlopes}.
We investigated what is the limiting precision of $\alpha$ needed to constrain the intrinsic scatter at the same level with 12 galaxies by artificially scaling $\delta_\alpha$.
The estimate of that limiting precision is shown with the solid black line highlighting that at least 100 kinematic tracers per galaxy are needed for sufficient precision.
For galaxies with fewer observed GCs the modelling has to be supplemented by stellar kinematics 
 or other halo tracers, such as planetary nebulae, to provide sufficient constraints on the $\alpha$ to recover the intrinsic scatter.
Increasing the number of galaxies will have an impact on the recoverability of the $\sigma_\alpha$ but more detailed quantification of the observing guidelines is beyond the scope of this work.

\subsection{The relation between the inner slope and stellar mass}\label{ch:RL_gamma_tot_Lit_compariosn_Global}

\begin{figure*}
    \centering
    \includegraphics[width = \textwidth]{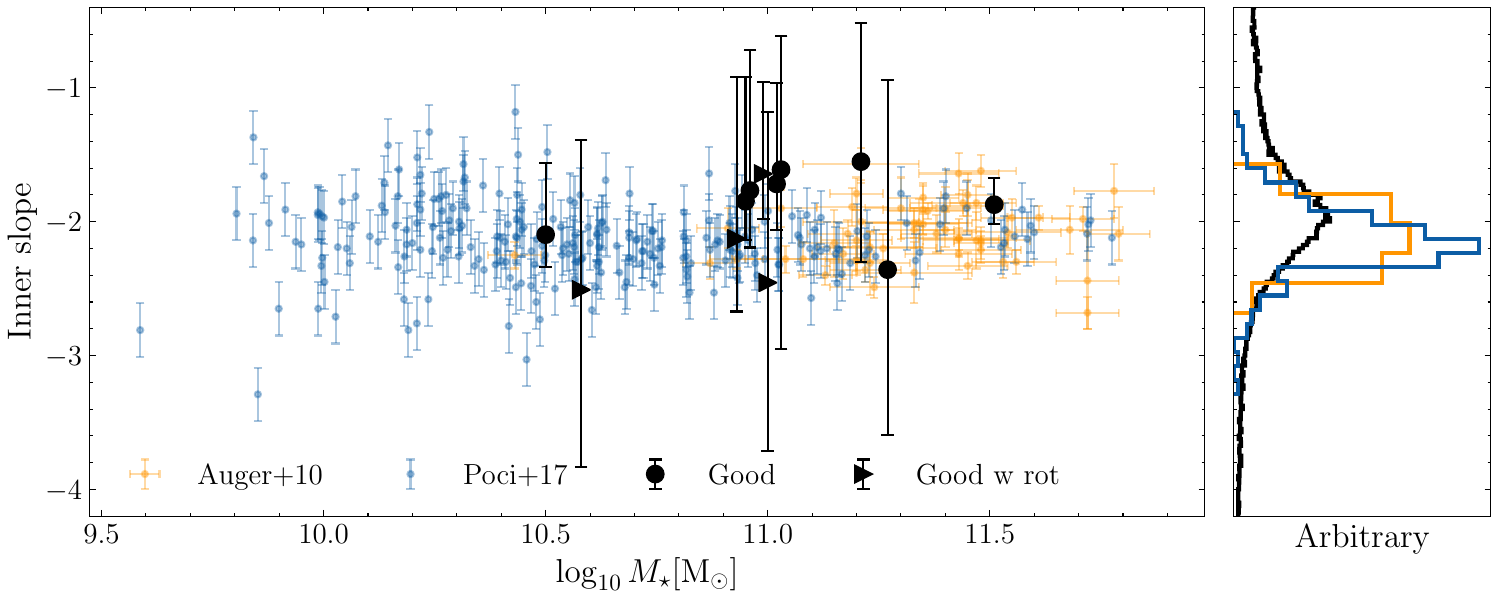}
    \caption{Relation between the stellar mass and inner slope on the left panel and the distribution of the slopes as determined by this work and literature samples in the right panel. Measurements from this work show the intrinsic inner slope - $\alpha$ of a generalised NFW and literature shows the single power-law slope - $\gamma_{tot}$.
    The black-filled circles and triangles show the galaxies with robust measurements representing the inner mass slope $\alpha$ as measured in this study from the GC kinematics, extending out to $\sim$8~\Reff.
    Symbols correspond to those of Fig.~\ref{fig:EnclosedMass_prior}.
    Large literature samples focused on the single power-law slope only. 
    They are shown as blue and orange points from \cite{Poci2017_TotalMassSlopes} and \cite{Auger2010} respectively. The gravitational lensing from the former study constrained the slope within 0.5~\Reff and the latter used stellar kinematics within $\sim 0.9$~\Reff.
    The arbitrary scaled distribution of the measurements on the right panel shows the distribution of the slopes with colours corresponding to that of the left panel. 
    }
    \label{fig:Gamma_StellarMass}
\end{figure*}

Several relations have been reported in the literature linking $\gamma_{\rm tot}$ to different properties of the galaxy.
Due to the limited number of galaxies with robust measurements, we focus only on the relation between the stellar mass and inner slope and present our measurements of the intrinsic inner slope $\alpha$ shown in Fig.~\ref{fig:Gamma_StellarMass}.
The literature values in light blue from \citetalias{Poci2017_TotalMassSlopes} were computed with the central stellar kinematics with median radius probed 0.9~\Reff.
Similarly, the orange points show measurements with gravitational lensing \citep{Auger2010} with median radius probed 0.5~\Reff.

We do not observe a statistically significant trend of $\alpha$ with the stellar mass. 
Similar has been reported by \citetalias{Bellstedt2018_EnclosedMassProfile}   and has also been found in the Magneticum simulations\footnote{www.magneticum.org} \citep{remus+17_fDM_vs_gammaTOT}.
\cite{Tortora+14_Central_slopes_Massive_gals} used spherical isotropic Jeans equations to carry out dynamical modelling of  SPIDER galaxies over large mass range and found the $\gamma_{\rm tot}$ become progressively shallower with increasing stellar mass.
A trend not observed by our study or by \cite{Cappellari+15_StellarKin_InnerSlope}.

The large spatial extent to which the discrete dynamical modelling of GCS is sensitive means that we have tracers beyond the scale radius of the total mass profile.
Therefore, the SLUGGS GCs are sensitive to the intrinsic inner total mass slope $\alpha$ and we can compare $\alpha$ to the values from the power-law studies of the inner mass slope but its interpretation is not straightforward.

\section{Discussion}\label{ch:Discussion}

Comparing with the literature results from \citetalias{Alabi2016_EnclosedMass}, \citetalias{Poci2017_TotalMassSlopes}
and \citetalias{Bellstedt2018_EnclosedMassProfile} we investigated various properties of the galaxies for potential trends with the environment (field, group, cluster), baryonic content (stellar mass), $(v_{\rm rot}/\sigma)_{\mathrm{\rm GCS}}$ as reported by \citetalias{Alabi2016_EnclosedMass}, flattening of the GCSs and a number of dynamical tracers on $\Delta\gamma_{\rm tot}$. 
We did not identify any parameters correlated with systematic offset for any of the 3 literature datasets. 

\subsection{The link between inner slope and galaxy formation}\label{ch:discussion_alpha_gal_formation}

The apparent universality of the $\gamma_{\rm tot}\sim-2$ has been linked to the formation history of the ETGs.
\cite{Remus_13} found in merger simulations that dry major mergers redistribute baryonic and dark matter such that the resulting total mass distribution follows a singular isothermal sphere.
This configuration is very stable and only the accretion of large amounts of gas can modify the inner slope.
Major mergers were more frequent in the earliest phases of galaxy cluster assembly and \cite{Derkenne+21_FF_inner_slope, Derkenne+23_MAGPI_inner_slope} recently studied redshift evolution of $\gamma_{\rm tot}$ finding shallower profiles in cluster galaxies compared to those in intermediate density environments at $z\sim0.3$.
With the trend persisting to $z=0$ and galaxies in intermediate density environments, that experience fewer major mergers, showing negligible evolution of  $\gamma_{\rm tot}$ in the past 3-4~Gyrs.
Recently, \cite{Zhu_K+23_Manga} showed a strong dependence of the inner slope on the $\sigma_{\rm e}$ and morphological type.

Different galaxy simulations disagree on the values for the inner slope. 
\citetalias{Bellstedt2018_EnclosedMassProfile} found that EAGLE\footnote{Evolution and Assembly of GaLaxies and their Environments.} simulations \citep{Schaye_15_EAGLE} produce galaxies with profiles shallower by $\sim0.3-0.5$ than Magneticum simulations \citep{remus+17_fDM_vs_gammaTOT}.
Galaxies in the latter simulation additionally show anticorrelation between the stellar mass and the inner slope $\gamma_{\rm tot}$ that is not observed in real galaxies.
Moreover, \cite{remus+17_fDM_vs_gammaTOT} studied the correlations between the galaxy properties and $\gamma_{\rm tot}$ for three different simulations that vary in whether they incorporate active galactic nuclei (AGN) and star-formation feedback.
They find that the AGN feedback in their simulations acts to erase the correlation between the slope and stellar mass of the simulated ETGs.
When comparing the results for galaxies from the Magneticum simulation with those from the Oser simulations \citep{Oser_2012_simulations} that do not include AGN feedback, they find large discrepancies and systematically steeper values of $\gamma_{\rm tot}$ for Oser simulated ETGs. 
The study also finds evidence that the slope in simulated galaxies is steeper at higher redshift and a strong anti-correlation between the central stellar surface brightness and $\gamma_{\rm tot}$.

Literature investigations have repeatedly found a  narrow intrinsic spread in $\gamma_{\rm tot}$ 
 (with $\sigma_\gamma^{\rm int.}\sim0.16$) for ETGs (\citetalias{Poci2017_TotalMassSlopes}, \citealt{Auger2010}).
Two studies with extended potential tracers found an intrinsic spread of 0.11 for ETGs using stellar kinematics out to 4~\Reff\ \citep{Cappellari+15_StellarKin_InnerSlope} and coupling central stellar kinematics with HI circular velocity out to 6~\Reff\ \citep{Serra+16_stellar_gas_innerslope_scatter}.
Studies have found that this bulge-halo conspiracy requires a non-universal stellar initial-mass function and feedback mechanisms to counteract dark matter contraction resulting from dark matter cooling \citep[e.g.][]{Dutton+14_explaining_small_Scatter}.
However, observations have primarily relied on gravitational lensing or IFU\footnote{Integral field unit.}-based dynamical modelling.
Both provide information well within 1~\Reff, where dark matter fraction is on average $<15\%$.
While \cite{Auger2010} used a single power-law slope for the total mass density, \citetalias{Poci2017_TotalMassSlopes} had a more realistic double power-law profile with a fixed transition radius.
\citetalias{Bellstedt2018_EnclosedMassProfile} used extended stellar kinematics out to 4~\Reff. However, the spatial extent of the tracers was not sufficient to freely vary the scale radius of the total mass profile.

Recently, \cite{Derkenne+23_MAGPI_inner_slope} showed that fixing the scale radius has a systematic impact on the recovered slope $\alpha$.
We have also shown that the radial region where the inner slope $\gamma_{\rm tot}$ is fit will impact the recovered value of the parameter as seen in \autoref{tab:gamma_tot_results}, a trend commented also by \citetalias{Bellstedt2018_EnclosedMassProfile}.
With discrete dynamical modelling of the GCs, we have shown in this work that it is possible to recover the intrinsic inner slope of the total mass profile, relaxing the assumption on the scale radius.
This opens an avenue for dynamical studies of a larger number of galaxies to probe the intrinsic slope $\alpha$ and better understand the interplay between the baryonic and dark matter physics in the process of galaxy formation.

In this work, we agree with the literature studies and find the mean inner slope $\gamma_{\rm tot}$ is steeper than isothermal, but the exact value of the profile is strongly dependent on the radial range where the profile is fit. 
Additionally, because of our flexible modelling approach and the large spatial extent of our tracers, our results are sensitive to the intrinsic inner slope of the double power-law profile $\alpha$.
As can be seen on Fig.~\ref{fig:Total_Mass_distribution_3115} the projected positions of the GCs in NGC~3115 can also be found around the best-fit scale radius of the total mass profile.
Meaning that there is enough power in the data to constrain the intrinsic inner slope.

\subsection{Limitations of our modeling approach}\label{ch:Subpop_Other_lit_comparison}

Discrete dynamical modelling is very flexible, yet great care is needed to understand the intrinsic and observational properties of the GCs.
Kinematically cold substructures or biased sampling of the underlying velocity field can result in the biased inference of the modelled parameters, an effect that is strongly dependent on the number of the kinematic tracers.

\subsubsection{Effect of subpopulations}\label{ch:Results_effect_of_subpop}

A subset of galaxies shows the presence of strong subpopulations that are not relaxed or trace the group/cluster potential instead of the galaxy potential.
These galaxies were identified by the rising or constant velocity dispersion out to 7~\Reff\ from the centre without a strong presence of rotation.
NGC~1407, NGC~4486 (M~87) and NGC~5846 are all the central galaxies of their group and NGC~4486 is the second brightest galaxy of the Virgo cluster.
Their GCS have been extensively studied in the literature.

\cite{Zhu16_5846} and \cite{Napolitano+14_5846} showed how the velocity dispersion of red GCs in NGC~5846 rises in the outskirts of the galaxy.
Both red and blue GCs also show distinct orbital anisotropy with blue showing mild radial anisotropy and red tangential anisotropy in the outskirts.
Similarly, \cite{Pota+15_NGC1407} found a kinematically distinct red and blue population of GCs in NGC~1407, with the velocity dispersion of blue GCs increasing towards the outskirts.
\cite{Strader+11_NGC4486} comprehensive kinematic and dynamical study M~87 (NGC~4486) revealed complexity and strong presence of substructures in the GCS of the galaxy pointing to active assembly.

To investigate whether the presence of subpopulations is causing the systematic offsets we modeled separately the blue and red populations of GCS in NGC~1407 as a test case.
\cite{Romanowsky2008_NGC1407} carried out a kinematic analysis of the red and blue populations using the colour information $(g'-i') = 0.98$ to separate the populations.
They found that the dispersion profile decreases until $R\sim20$~kpc and then starts to increase. 
They report low $(v_{rot}/\sigma)_{\rm GCS}$ in the outer parts and find signs of misaligned rotation signature between the red and blue population with hints of tangentially biased orbits in the outer regions. 
The blue population shows a presence of a cold moving group around $R\sim40$~kpc.

Following the work by \cite{Romanowsky2008_NGC1407} we separate the two populations of GCs based on their colour.
Then carry out the same completeness and tracer density analysis as described in Sect.~\ref{ch:TD_profile} and re-run the dynamical modelling.
We find that the red population, which is more centrally concentrated and shows declining velocity dispersion in the outer parts, results in a steeper inner slope, in agreement with the literature results.
The dispersion profile of the blue population stays flat in the entire radial range and the resulting inner slope is shallower than that obtained for the red GCs.
The enclosed mass measurements from the red population are also in excellent agreement with the literature, while the blue populations result in overestimating the mass by several orders of magnitude.
We aim to investigate the other two galaxies and present the results for NGC~1407 in a follow-up study.

\subsubsection{Peculiar GCS}\label{ch:Results_effect_of_peculiar}

A subset of galaxies, NGC~4374, NGC~4459, NGC~4526, and NGC~7457 showed peculiar trends in their velocity field.
All but NGC~4526 have fewer than 40 GCs sparsely populating their halo. 
NGC~4526 has a peculiar velocity field in the GCS with the central region showing strong rotation and the outer region showing a velocity offset. 
This particular galaxy has twice the number of GCs compared with the other galaxies flagged as Subpop. and the recovered enclosed mass is consistent with the literature studies by \citetalias{Alabi2016_EnclosedMass} as highlighted in Fig.~\ref{fig:EnclosedMass_5Re}.
However, the inner slope is more sensitive to the peculiar features in the observed GC velocity field, resulting in a biased measurement of the inner slope $\alpha$ shown on Fig.~\ref{fig:BellSLope}.

To understand the discrepancy with the literature studies we explored trends with the galaxy stellar mass, environment, and rotation parameter as reported by \citetalias{Alabi2016_EnclosedMass}, the \sersic index and the number of tracers.
We found the latter has the strongest impact together with a non-uniform spatial distribution of the tracers.
These observational effects result in the shallower profile and overestimated mass when using discrete dynamical modelling. 
Similar to the galaxies discussed in Sect.~\ref{ch:Results_effect_of_subpop} we consider our measurements for galaxies flagged as Subpop. are not robust.

\section{Summary and future work}\label{ch:Conslussions}

We carried out the discrete dynamical modelling of a large number of GCSs from the SLUGGS survey.
Detailed analysis of the tracer density profile and the investigation of the GC velocity field limited the number of galaxies we modelled from 27 to 21 and we present robust measurements for 12 galaxies.
These galaxies have a sufficiently large number of kinematic tracers without a strong observational footprint or presence of subpopulations.
We used Bayesian inference and MCMC sampler to efficiently sample the posterior and find the best-fit parameters.
We find that informative priors on  $\log\rho_{\rm s}$ and $\log r_{\rm s}$ can help break the degeneracy in these parameters.

We explore 3 different model setups (A, B, C) to investigate how the modelling assumptions impact the best-fit parameters.
Models A and B assume non-rotating and isotropic GCs and model C explores the impact of velocity anisotropy and rotation on the recovered parameters of the total mass profile.
We find that $\beta_z$ and $\kappa$ have a negligible impact on the enclosed mass and inner slope.
We identify model B as the fiducial model and find the logarithmic inner slope of the 12 galaxies with robust measurements is $\langle \gamma_{\rm tot}\rangle = -2.22\pm0.14$ in the range 0.1-4~\Reff.
These values agree within the uncertainties but we find that our values are systematically shallower than the literature measurements.
This could be a result of our more flexible modelling approach.
The large spatial extent of the kinematic tracers in tandem with the discrete modelling approach enables us to constrain the transition radius of the total mass slope and measure the intrinsic inner slope $\overline{\alpha} = 1.88\pm 0.01$ that is shallower than $\langle \gamma_{\rm tot}\rangle$.

We find a large rms scatter in the inner slope which is in part driven by the large measurement uncertainties.
We demonstrate that in order to constrain the intrinsic scatter at the level of 0.16 at least 100 kinematic tracers are needed per galaxy.

In Sect.~\ref{ch:Results_effect_of_subpop} we have shown that the modelling results of dispersion-dominated systems with \textsc{cjam} are minimally affected by rotation and anisotropy when the number of tracers is sufficiently large, but providing quantitative guidelines is beyond the scope of this work.
When modelling merger remnants and GCS with a strong presence of distinct subpopulations, our results are strongly biased towards overestimating the enclosed mass by several orders of magnitude and shallower profile.

Both simulations and observations provide strong evidence that the total mass slope cannot be extrapolated with a single power-law profile at a large spatial extent and new observations will enable both larger photometric and kinematic samples of discrete tracers in the nearby universe.
We can use them to characterize the mass distribution on large scales and constrain structure formation models.
This work highlights the benefits and caveats of using sparse dynamical tracers in the haloes of the galaxies.
Combining integrated stellar kinematics in the inner regions with more extended halo tracers can provide stronger constraints on the central mass distribution.
Our work can be expanded by including higher Gauss-Hermite moments or assuming an intrinsic 3D Gaussian velocity distribution instead of the 1D projected.

With the upcoming wide-field photometric surveys larger samples of extragalactic GCs will be observed and this will open an avenue for more studies of these low surface-brightness diffuse haloes that contain rich information on the total mass content of galaxies.
With the full mass profiles, dark matter content can also be investigated over large galactocentric radii. 
We would also like to further explore methods to improve the precision of the modelling results and better account for the kinematic substructures.

\begin{acknowledgements}
We thank the referee for the helpful comments and suggestions that have improved this manuscript.
TV thanks Marina Rejkuba, Lucas Valenzuela and Rhea Silvia Remus for their feedback and discussion on the manuscript.
This project has received funding from the European Research Council (ERC) under the European Union’s Horizon 2020 research and innovation programme under grant agreement No 724857 (Consolidator Grant ArcheoDyn). TV acknowledges the studentship support from the European Southern Observatory. 
The computational results presented have been achieved (in part) using the Vienna Scientific Cluster (VSC).

This work made use of the following software:
\textsc{NumPy} \citep{Numpy2020}, \textsc{SciPy} \citep{Scipy}, \textsc{Matplotlib} \citep{Matplotlib}, \textsc{Astropy} \citep{astropy:2013, astropy:2018, astropy:2022}.

\end{acknowledgements}

\bibliographystyle{aa}
\bibliography{biblio.bib}

\end{document}